\documentclass[%
 reprint,
 amsmath,amssymb,
 aps,
prx,
nolongbibliography,
]{revtex4-2}

\usepackage{dcolumn}
\usepackage{bm}

\usepackage{graphicx}%
\usepackage[normalem]{ulem}
\usepackage{xcolor}%
\usepackage{algorithm}%
\usepackage{algorithmicx}%
\usepackage{algpseudocode}%
\usepackage{listings}%

\lstdefinelanguage{LAMMPS}{
    keywords={units, atom_style, pair_style, pair_coeff, read_data, fix, run, timestep, velocity, thermo, group, region, create_atoms, variable, equal,compute, norm, ave},
    sensitive=true,
    comment=[l]\#,
    morestring=[b]",
}

\usepackage{comment}
\usepackage{anyfontsize} 

\usepackage{siunitx}
\usepackage{chemmacros}

\usepackage{booktabs} 

\usepackage{enumitem}

\setlist[itemize]{leftmargin=1.8em, itemsep=0pt, parsep=0pt, topsep=0pt}      
\setlist[enumerate]{leftmargin=1.8em, itemsep=0pt, parsep=0pt, topsep=0pt}  

\renewcommand{\sout}[1]{\unskip}

\begin{document}

\title{Critical surface phase behavior governs hydrophobic attraction between extended solutes}


\author{Nigel B. Wilding}
 \email{nigel.wilding@bristol.ac.uk}
\author{Francesco Turci}%
 \email{f.turci@bristol.ac.uk}
\affiliation{H.H. Wills Physics Laboratory, University of Bristol, Royal Fort, Bristol BS8 1TL, UK}

\begin{abstract}

Hydrophobic interactions are central to biological self-assembly and soft matter organization, yet their microscopic origins remain debated. A key hallmark is the strengthening of attraction between hydrophobic solutes with increasing temperature, a feature often attributed to entropy changes from disrupted hydrogen bonding in water. Here we present an alternative framework based on surface phase behavior, supported by extensive molecular dynamics simulations. Using metadynamics, we quantify the solvent-mediated effective potential between nanometer-scale hydrophobic solutes in the monatomic water (mW) model, the SPC/E water model, and a Lennard-Jones solvent. We develop a morphometric model which incorporates a scaling theory of critical drying, linking the range and strength of hydrophobic attraction to interfacial thermodynamics and proximity to vapor–liquid coexistence. The model reproduces the effective potential across diverse solute sizes, degrees of hydrophobicity, and thermodynamic states. Our simulations recover the inverse temperature dependence of hydrophobicity, showing it arises generically from rapid thermal expansion of the solvation shell.

\end{abstract}
\maketitle

\section{Introduction}
\label{sec:intro}

Water is undoubtedly the most important solvent on Earth, serving a key role in numerous biological and chemical processes. While many solutes dissolve readily in water, hydrophobic solutes -- typically large, nonpolar molecules -- tend to separate from water, minimizing direct interaction. Amphiphilic molecules, containing both hydrophilic and hydrophobic components, exhibit similar behavior by forming micelles in aqueous environments. 
These phenomena exemplify the ``hydrophobic effect'', wherein hydrophobic entities experience water-mediated attraction. 

Extensive experimental and simulational studies of hydrophobic solutes in water~\cite{Rein_ten_Wolde:2002,Southall:2000aa,Widom2003,Naito2003,Huang:2003aa,Bagchi:2005aa,Bowron:2007aa,Sobolewski:2007aa,Tomlinson-Phillips:2011aa,Maiti:2012aa,Chaudhari2013,Koga:2013aa,Dias:2014aa,Bischofberger:2014aa,Ben-amotz2016,Pratt:2016aa,Gao:2018aa,Koga:2018aa,Conti-Nibali:2020aa,Bogunia:2022aa,SUN2022111550,SUN2019199,Havenith2022,Ghosh:2023,Naito2024} have explored the factors influencing the hydrophobic effect. These studies highlight the importance and interplay of various factors such as the degree of solute hydrophobicity, the solute size, pressure, and temperature.  
One particularly counterintuitive feature of the effect is that the strength of the attraction between solutes typically increases with rising temperature~\cite{Pratt:2016aa}. Experimentally, this is observed in apolar polymer coils, which demonstrate a lower critical solution temperature in water, remaining mixed at lower temperatures but separating as the temperature rises~\cite{Davies:2000aa,Moelbert:2003aa,Xie2022}. Another manifestation is the temperature dependence of solubility which decreases with rising temperature~\cite{DillBromberg,Widom2003}. 

This `inverse temperature effect' is often attributed to water's unique properties. Based on a suggestion first proposed 80 years ago~\cite{FrankEvans1945}, it is widely considered that a ``hydration shell'' surrounds hydrophobic solutes, a structured layer of hydrogen-bonded water molecules more ordered than the bulk water~\cite{Silverstein:2000aa,Lee:1996aa,Widom2003,Paschek:2004,Raschke2005,Zangi2008_8634,Bischofberger:2014aa,GrdadolnikAvbelj2016,GrabowskaZielkiewicz2021}. As temperature increases, this ordered structure breaks down, leading -- it is often argued -- to a rise in entropy that promotes solute aggregation ~\cite{Silverstein:2000aa,Widom2003,Paschek:2004,israelachvili2011,Bischofberger:2014aa,KRONBERG201614,Ghosh:2023}. However, clear experimental evidence, eg. from neutron scattering studies, for the link between changes in hydrogen-bonded structures near a solute and the subtle thermodynamics of the hydrophobic effect seems lacking~\cite{Soper1993,Soper:2006aa} and the matter remains controversial, see e.g. \cite{Ball_water_2008,Ben-Naim2013,Kim:2015aa,GrabowskaZielkiewicz2021}.

A different perspective on hydrophobic solvation phenomena is provided by the theory of Lum, Chandler, and Weeks (LCW)~\cite{LumChandlerWeeks1999,HuangChandler2000PNAS,Chandler:2005aa}, which focuses on the number density profile of water near a solute.  LCW theory offers a seminal microscopic framework, highlighting the emergence of a layer of depleted density around an isolated extended solute. It emphasizes that hydrophobic behavior differs for small solutes, which can integrate into the hydrogen bond network, compared to larger solutes that disrupt this network, creating a distinct vapor-liquid-like interface where surface tension rather than hydrogen bonding plays a key role. LCW theory has been successful in explaining several key aspects of hydrophobicity. However, as we shall show, by incorporating the role of fluctuation-driven phenomena in descriptions of hydrophobic solvation, one can form a more comprehensive and interpretable understanding of the key phenomenology.

The present work considers the case of extended solutes with a radius of a nanometer or greater, relevant to large molecules, proteins, and nanoparticles, where a detailed microscopic understanding of hydrophobic attraction and its dependence on factors such as solute size, degree of hydrophobicity, pressure, and temperature remains elusive.Recent studies on extended spherical hydrophobic solutes have linked solvation phenomena to the physics of the critical drying transition. This surface phase transition occurs when water at bulk vapor-liquid coexistence interacts with a superhydrophobic planar surface~\cite{CoeEvansWildingPRL2022,Coe2023,EvansWilding2015,EvansStewartWilding2016,EvansStewartWilding2017,EvansStewartWilding2019}. By analyzing radial density and compressibility profiles around model solutes, these studies have revealed detailed physical features of the solvation shell where the presence of the solute significantly alters local water properties.  A bespoke scaling theory elucidates how these characteristics depend on factors such as the solute radius, the strength of solute-water attraction, and external thermodynamic fields describing the proximity to bulk vapor-liquid coexistence.

We build on these new insights by linking the properties of hydrophobic solvation to the effective water-mediated attractive radial potential, $W(r)$, between a pair of identical spherical hydrophobes. We propose a minimal morphometric model that describes this potential based on the geometry and physical properties of overlapping solvation shells of elevated free energy, and integrating key insights derived from a scaling description of hydrophobic solvation. The model’s predictions are compared with extensive molecular dynamics (MD) simulations of $W(r)$, demonstrating a quantitative match with the detailed shape of the potential and clarifying its dependence on solute radii, pressure, temperature, and the strength of solute-water attraction. Specifically, we present evidence that the well-known increase in hydrophobic attraction with rising temperature stems from a previously overlooked phenomenon: a significant expansion of the solvation shell surrounding a solute, a feature associated with critical drying. Additionally, we show that this inverse temperature effect also manifests for a pair of extended solvophobic solutes in a Lennard-Jones solvent, demonstrating that hydrogen bonding is not a prerequisite for the effect. A short report of some aspects of this work has previously appeared elsewhere~\cite{WildingTurciPRR2025}.
    
\section{Hydrophobic solvation and critical drying}
\label{sec:solvation}

\subsection{Background}

Simulation studies of water models near extended hydrophobic spherical solutes reveal a layer of depleted water density accompanied by enhanced density fluctuations~\cite{LumChandlerWeeks1999,Huang:2000wq,Sarupria2009,AcharyaGarde2010,MamatkulovKhabibullaev2004,PatelVarillyChandler2010,Mittal:2008aa, Oleinikova:2012aa,VaikuntanathanE2224,Patelreview2022}. Coe {\em et al.}~\cite{CoeEvansWildingPRL2022,Coe2023,CoeThesis} performed a comprehensive study of these features using simulation and density functional theory, attributing key aspects of their behavior to the proximity of a critical drying surface phase transition. This transition occurs when a solvent is (i) at vapor-liquid coexistence and (ii) in contact with a strongly solvophobic {\em planar} substrate ~\cite{EvansStewart2015,EvansStewartWilding2017,EvansStewartWilding2019}. As the transition is approached, e.g., by lowering the liquid phase pressure ($p$) to its coexistence value $p_{coex}$, a vapor layer intrudes between the substrate and the bulk liquid, and the contact angle approaches $180^\circ$. The width of this vapor layer diverges smoothly at the transition point, as does the local compressibility, which reflects the strength of density fluctuations at the emerging vapor-liquid interface. The corresponding critical exponents are known~\cite{EvansStewartWilding2016}.

Under ambient conditions, water in contact with a finite spherical hydrophobe is not at a critical drying transition. However, ambient water is {\em close} to coexistence, with the degree of supersaturation being very small (the coexistence pressure is $p_{coex}\approx 1/20$ atm). Consequently, the free energy cost of creating vapor regions within the stable liquid is low~\cite{LumChandlerWeeks1999}. Coe {\em et al.} developed an effective interface (``binding potential") theory for extended spherical solutes that describes the near-critical scaling of the vapor layer width and the strength of enhanced density fluctuations with the solute radius, the degree of oversaturation, and the strength of solute-water attraction.   The scaling predictions were validated through extensive mW simulations and classical density functional theory calculations within the grand canonical ensemble~\cite{CoeEvansWildingPRL2022,Coe2023,bui_classical_2024}, down to surprisingly small values of the solute radius $R_s\approx 10${\AA}.

\subsection{Radial density and compressibility profiles for a hydrophobic solute in water}

To establish that signatures of near-critical drying appear at the length scales characteristic of large molecules, we present new results from molecular dynamics (MD) simulations performed in the isothermal-isobaric (constant-$NpT$) ensemble. The simulations, described in detail in Appendix~\ref{app:mD}, utilize the widely adopted coarse-grained monatomic water (mW) model~\cite{MolineroMoore2009}, outlined in Appendix~\ref{app:mW}, whose vapor-liquid phase diagram and surface tension have been extensively characterized~\cite{CoeEvansWildingJCP2022}. The mW model effectively captures water's key structural and energetic properties, including its tetrahedral network and thermodynamic anomalies. Our study focuses on water in contact with an isolated, purely repulsive (and therefore maximally hydrophobic) spherical solute particle, with solute interactions described by a Morse potential as detailed in Appendix~\ref{app:interpot}. We analyze the behavior of three types of radial profiles to expose the near-critical behavior:

\begin{enumerate}

\item {\bf The average number density profile of water molecules}. This is defined as the ensemble average
\begin{equation}
\varrho(r)\equiv\langle\rho(r)\rangle,
\end{equation}
of the instantaneous density profile  $\rho(r)=N(r)/4\pi r^2dr$, where $N(r)$ is the count of water molecules in a spherical shell extending from $r$ to $r+dr$.

\item 
{\bf The local compressibility profile}, a response function that quantifies the strength of local density fluctuations, providing insight into the hydro/solvophobicity of a solute. In the constant-$NpT$ ensemble, it is defined as:

\begin{equation}
\chi_p(r) \equiv \left.\frac{\partial\varrho(r)}{\partial p}\right|_T.
\label{eq:chidep}
\end{equation}
  $\chi_p(r)$ is closely related to $\chi_\mu(r)$ (grand canonical ensemble)~\cite{EvansStewart2015,Wilding:2024} -- as well as other recently introduced fluctuation measures~\cite{EvansStewart2015,EckertSchmidt2020,CoeEvansWildingPRE2022,Eckert_2023}, though its use in the isobaric ensemble has not been previously examined.

In simulations, $\chi_p(r)$ is calculated from the covariance of the instantaneous density profile $\rho(r)$ and the fluctuating system volume $V$, (Appendix~\ref{app:chip}):

\begin{equation}
\chi_p(r) = -(k_B T)^{-1} \left(\langle \rho(r) V \rangle - \langle \rho(r) \rangle \langle V \rangle\right).
\label{eq:chicovar}
\end{equation}

\item {\bf The pressure derivative of $\chi_p(r)$}:
\begin{equation}
    \upsilon_p(r)\equiv\left. \frac{\partial^2 \varrho(r) }{\partial p^2}\right|_T.
    \label{eq:upsilondef}
\end{equation}
This quantifies the rate at which local density fluctuations grow as the system approaches vapor-liquid coexistence, as well as the local character of that growth. Similarly to $\chi_p(r)$, it can be expressed in terms of covariances, though for numerical reasons we do not take this route here (see Appendix~\ref{app:chip} for further discussion). 

\end{enumerate}

\subsection{Near-critical fluctuations: dependence on pressure and solute radius}

To investigate the impact of water's proximity to the vapor-liquid coexistence curve on solvation properties, we examined $\varrho(r)$, $\chi_p(r)$, and $\upsilon_p(r)$ at pressures ranging from $0$ to $3000$ atmospheres (atm) while maintaining a constant temperature of $T = 300\,\text{K}$.

\begin{figure}
\centering
\includegraphics{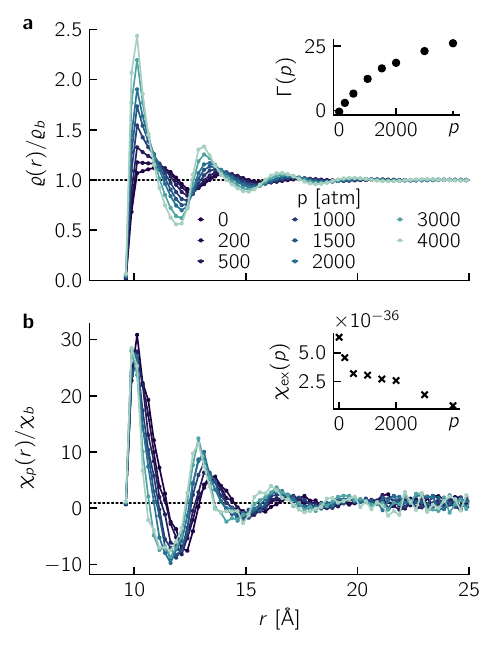}
\caption{Pressure dependence of radial profiles for mW water in contact with a purely repulsive solute of radius $R_s=10\,${\AA} at $T=300\,$K. (a)  Number density profile $\varrho(r)$, of water normalized by the bulk number density $\varrho_b$; the inset shows the $p$ dependence of the dimensionless total adsorption $\Gamma$, defined as in the main text. (b)  Compressibility profile $\chi_p(r)$, normalized by its bulk value $\chi_b = \varrho_b \kappa_T$, where $\kappa_T$ is the bulk isothermal compressibility; the inset displays the $p$ dependence of the total surface excess compressibility $\chi_{ex}$, defined in the main text,   which has SI units $J^{-1}m^3$. $\chi_p(r)$ is calculated using Eq.\,(\ref{eq:chicovar}). Bulk properties of mW are listed in Appendix~\ref{app:bulkprops}. 
}
\label{fig:rho-chi-Rs10}
\end{figure}

\begin{figure}
\centering
\includegraphics{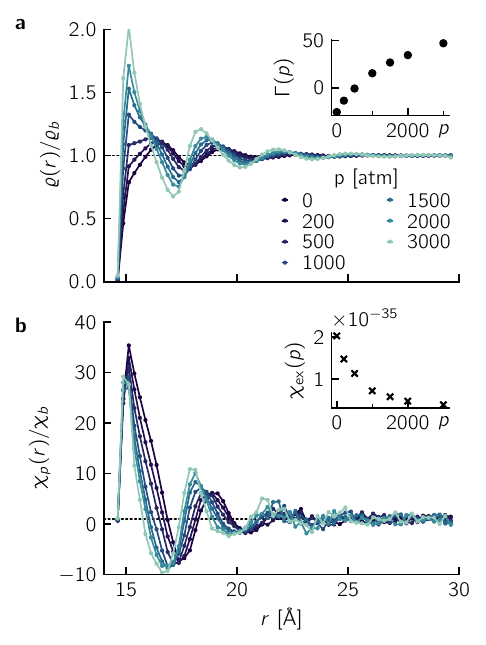}
\caption{As for Fig.~\ref{fig:rho-chi-Rs10} but for $R_s=15\,${\AA}. Note that $\Gamma$ is negative at coexistence ($p=0$), indicating pronounced depletion of density for this $R_s$.}
\label{fig:rho-chi-Rs15}
\end{figure}

Figure~\ref{fig:rho-chi-Rs10}(a) displays the spherically averaged density profile, $\varrho(r)$, for a single solute of radius $R_s = 10\,\text{\AA}$; pressure values are indicated in the legend. Each profile is normalized by the bulk number density $\varrho_b(p)$ at the respective pressure. Similarly, Fig.~\ref{fig:rho-chi-Rs10}(b) shows the local compressibility, $\chi_p(r)$, obtained using the covariance relation Eq.\,(\ref{eq:chicovar}) and normalized by the bulk compressibility, $\chi_b(p)$. For comparison, Fig.~\ref{fig:rho-chi-Rs15} presents the normalized density, $\varrho(r)/\varrho_b$, and local compressibility, $\chi_p(r)/\chi_b$, profiles for a larger solute of radius $R_s = 15\,\text{\AA}$.   Bulk properties of mW at the studied state points are listed in Appendix~\ref{app:bulkprops}.

The density profiles for the two solute radii in Figs. \ref{fig:rho-chi-Rs10} and \ref{fig:rho-chi-Rs15} show both common features and distinct differences. In all cases, $\varrho(r)$ exhibits oscillations from particle layering near the solute surface, similar to liquids near planar hard walls \cite{HansenMcDonald}, with the strongest oscillations at high pressure. As $p \to p_{coex}^+$ ($p_{coex} \approx 0$ atm on the scale relevant to $\varrho(r)$ and $\chi_p(r)$), both the oscillation amplitude and contact density decrease markedly, accompanied by a strong reduction in the dimensionless total surface adsorption $\Gamma \equiv 4\pi \int r^2 dr\, [\varrho(r) - \varrho_b]$, shown in the insets of Figs. \ref{fig:rho-chi-Rs10}(a) and \ref{fig:rho-chi-Rs15}(a).

The local compressibility profile $ \chi_p(r)$ shows a peak at the location of the developing vapor-liquid interface, reaching approximately 30 times the bulk value. Oscillations due to layering are observed at all pressures and follow the same pattern as $\varrho(r)$. As $p\to p_{coex}^+$, the width of the first peak in $ \chi_p(r) $ significantly broadens, and its height increases. While these signs of growing density fluctuations may initially seem subtle, this is because layering effects at these solute sizes partially obscure the emergent behavior. A more effective measure to capture the intrinsic pressure dependence is the total surface excess compressibility, $ \chi_{\text{ex}} \equiv 4\pi \int r^2 dr \, (\chi_p(r) - \chi_b) $, which has dimensions of $1/p$,  and is shown in the insets of Figs.~\ref{fig:rho-chi-Rs10}(b) and \ref{fig:rho-chi-Rs15}(b). This quantity sharply increases as $p \to p_{\text{coex}}^+ $, with the rate of increase accelerating closer to the coexistence point.

\begin{figure}
\centering
\includegraphics{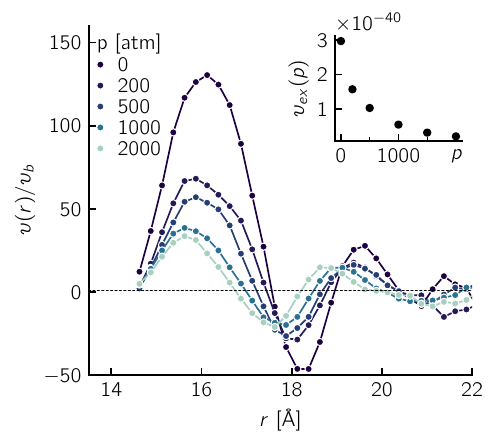}
\caption{ Measured forms of $\upsilon_p(r)$ obtained as a numerical pressure derivative of the profiles of Fig.~\ref{fig:rho-chi-Rs15}(b), as described in the main text. The inset shows the pressure dependence of the surface excess quantity $\upsilon_{\text{ex}} = 4\pi \int r^2 dr \, (\upsilon_p(r) - \upsilon_b)$,  which has units $J^{-2}m^6$.}
\label{fig:nu-Rs15}
\end{figure}

$\upsilon_p(r)$ [Eq. (\ref{eq:upsilondef})], the pressure derivative of $\chi_p(r)$, captures near-critical density fluctuations as $p \to p_{coex}^+$ while reducing sensitivity to layering effects. We estimate $\upsilon_p(r) $ via the numerical derivative: $(p' - p)^{-1}(\chi_{p'}(r) - \chi_p(r)) $. The results, shown in Fig.~\ref{fig:nu-Rs15}, derive from the dataset of Fig.~\ref{fig:rho-chi-Rs15}(b), and clearly illustrate the accelerated growth of $ \chi_p(r) $ on the approach to coexistence, beyond what is evident from $ \chi_p(r) $ alone. This enhancement is further quantified in the inset of Fig.~\ref{fig:nu-Rs15}, which plots the surface excess quantity $ \upsilon_{\text{ex}} = 4\pi \int r^2 dr \, (\upsilon_p(r) - \upsilon_b)$ (which has dimensions of $1/p^2$). 

A comparison of the solute radius $R_s$ dependence of $\varrho(r)$ and $\chi_p(r)$ reveals that near and at $p = p_{\text{coex}}$, the density depletion adjacent to the solute surface is more pronounced for the larger solute. Furthermore, the surface excess compressibility $\chi_{\text{ex}}(p)$ indicates that, at a given pressure, the larger solute induces significantly stronger overall density fluctuations than the smaller one.

The observed density depletion near the solute surface, along with the increase in $\chi_{\text{ex}}(p)$ and $\nu_{\text{ex}}(p)$ as $p \to p_{\text{coex}}^+$, and as $R_s$ increases, are signatures of the approach to a critical drying transition.  For a solvent near vapor-liquid coexistence interacting with a strongly solvophobic solute particle, the density close to the solute approaches that of the bulk vapor phase as the solute radius $R_s$ becomes sufficiently large. Previous simulations of the mW model~\cite{CoeEvansWildingPRL2022,Coe2023} found this convergence occurs for $R_s \gtrsim 40$~\AA.

\subsection{Nature of the solvation shell for nanometer-sized solutes: vapor bubbles}

The present study examines solutes with radii $R_s = 10$\;\AA{} and $15$\;\AA{}, a range representative of large molecules and small proteins. At these scales, the fluid-structure near the solute surface is not vapor-like but inherently inhomogeneous, characterized by a dynamic interplay between dense, liquid-like regions and sub-nanoscale vapor bubbles that form spontaneously at the hydrophobic solute surface~\cite{CoeEvansWildingPRL2022,Coe2023}. Hence, the fluid properties within the solvation shell differ qualitatively from those of both the bulk vapor and liquid phases. The observed pressure-induced variations in $\varrho(r)$ and $\chi_p(r)$ are primarily driven by changes in the population and size of these bubbles, and the resulting density profiles $\varrho(r)$ can be interpreted as a weighted average of vapor-like and liquid-like density profiles~\footnote{In the context of critical phenomena, the liquid-like regions within the solvation shell contribute an analytical (non-scaling) background that can obscure the growth of near-critical density fluctuations. The higher-order derivative $\upsilon(r)$ suppresses this background more effectively than $\varrho(r)$ or $\chi_p(r)$.}. 

The vapor bubbles herald the first appearance of the incipient vapor phase at the solute’s surface. Their shape reflects the anisotropic correlation lengths parallel and normal to the interface, $\xi_\parallel$ and $\xi_\perp$, which are central to critical drying theory~\cite{EvansStewartWilding2016}. As the transition is approached, both lengths diverge, albeit with different critical exponents. The local compressibility obeys $\chi_p(z)\propto\xi_\parallel^{2}$, so $\chi_p(z)$ furnishes a direct probe of density-density correlations around the solute. By contrast, the perpendicular length grows only as $\xi_\perp\sim\sqrt{\ln(\xi_\parallel/\xi_b)}$, with $\xi_b$ the bulk vapor correlation length~\cite{EvansStewartWilding2016,Evans1989}. Owing to this logarithmic dependence, $\xi_\parallel\gg\xi_\perp$, yielding strongly \emph{anisotropic} critical fluctuations: laterally the vapor bubbles extend far more than they do in thickness.

Simulations support this scenario for both planar substrates~\cite{EvansStewartWilding2017} and extended spherical solutes (see Fig.~14 of Ref.~\cite{Coe2023} for a visual example). They reveal that vapor bubbles tend to spread laterally and remain flat against the surface. The central insight is that, near a solvophobic surface, the thickness of the drying layer perpendicular to the surface can be on the order of the solvent’s molecular length, yet still exhibits enhanced compressibility because the dominant density fluctuations occur parallel to the surface. 

The above arguments and observations motivate the interpretation of the properties of hydrophobic solvation for extended solutes in terms of scaling behaviour derived and adapted from the physics of critical drying.

\section{A coarse-grained model for hydrophobic attraction between spherical solutes}
\label{sec:model}

\begin{figure}
\centering
\includegraphics[width=.85\linewidth]{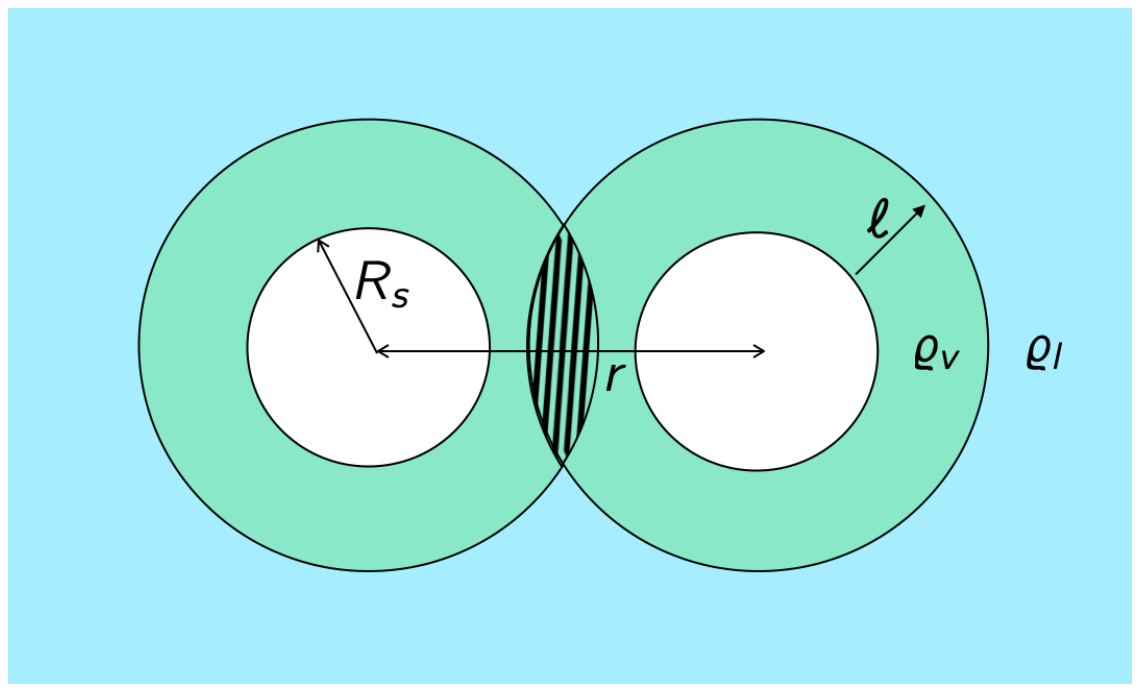}
\caption{
Schematic diagram (not to scale) of two equal hydrophobic solutes of radius $R_s$, each surrounded by a solvation shell of width $\ell$ (on the Å scale). The solute centers are separated by distance $r$, with the shells overlapping in the hatched region. The water number density in the solvation shell is $\varrho_v$, and in the surrounding bulk liquid is $\varrho_l$.
}
\label{fig:overlappingspheres}
\end{figure}

Solvent-mediated interactions are well studied in colloidal suspensions \cite{Grier:1997aa,Belloni2000,PhysRevE.65.061410,Malijevski:2015,Malijevski-Parry2015,LamLutsko2018} and, for hydrophobic systems, in water-mediated forces between nanoscale and macroscopic surfaces \cite{Zangi:2008aa,Giovambattista:2009aa,Kanduc:2016aa,LamLutsko2018,Riva:2024aa}, but less so for spherical solutes at intermediate scales. 

We use a minimal morphometric model~\cite{Bryk2003,Koenig2004} describing smooth overlap of solvation shells, conceptually related to depletion attraction~\cite{AO1954,poon_physics_2002} but including a liquid-vapor interface\footnote{Valid when first-order bridging transitions can be neglected~\cite{Malijevski:2015,Malijevski-Parry2015} for $R_s \lesssim 30$ Å in water.}. The shell is a homogeneous region of finite width where the free energy density exceeds that of the bulk, with an interface characterized by surface tension. This coarse-grained approach omits explicit vapor bubbles or molecular detail, incorporating their effects into model parameters. We then extend it to include the dependence of shell width on solute size, hydrophobicity, and thermodynamic state~\cite{Coe2023}. 

Initially, we consider two strongly hydrophobic spheres of radius $R_s$ in oversaturated water, with purely repulsive solute-solute and solute-water interactions, assuming $R_s$ is large enough that the shell is vapor-like~\cite{Coe2023}. Consider the evolution of the grand potential as the spheres move closer together and their solvation shells overlap, as schematically depicted in Fig.~\ref{fig:overlappingspheres}. The width of the hydrophobic solvation shell is denoted by $\ell$, and we define the combined radius of the solute and vapor layer as ${\mathcal R}=R_s+\ell$. The effective solute-solute potential, also known as the ``potential of mean force'', is expressed as

\begin{equation}
    W_m(r)=\Omega_{ex}(r)-\Omega_{ex}(r\to\infty)\:,
    \label{eq:Omegadef}
\end{equation}
where $\Omega_{ex}(r)$ is the excess grand potential of the system when the solute particles are separated by a linear distance $r$.

As $r$ is reduced in the range $2R_s\leq r\leq 2{\mathcal R}$ so that the solvation shells progressively overlap, we can identify two contributions to $W(r)$:

\begin{enumerate}

\item {\bf Interfacial area reduction:} 

Overlap of solvation shells reduces the total interfacial area to bulk liquid. For two intersecting spheres of radius $R_s$, the surface area is

\begin{equation}
A_u(r)=4\pi {\mathcal R}^2 +2\pi {\mathcal R} r\:.
\end{equation}

The reduction relative to $r \to \infty$ is $\delta A(r) = -2\pi (2\mathcal{R}^2 - \mathcal{R} r)$, giving a contribution $\delta A(r)\gamma_{lv}$ to $\Omega_{ex}(r)$, where $\gamma_{lv}$ is the vapor–liquid surface tension, which is a decreasing with $T$.

\item {\bf Increase in bulk volume}:  As the solvation shells overlap, the total volume available to the more stable liquid phase increases. The increase is the negative of the volume of the lens formed by two intersecting spheres of equal radii ${\mathcal R}=R_s+\ell$, separated by a distance $r$, for which one finds a cubic dependence on $r$

\begin{equation}
V_o(r) =\frac{\pi}{12}(4{\mathcal R}+r)(2{\mathcal R}-r)^2   \:.
\end{equation}
The corresponding contribution to $\Omega_{ex}(r)$ is $-V_o\delta p$ where $\delta p$ is the local difference in the grand potential free energy density between the bulk liquid and the solvation shell, which for a true vapor solvation layer is equal to the oversaturation $(p_l-p_v)$. 

\end{enumerate}
Inserting these results into Eq.\,(\ref{eq:Omegadef}) and setting $\Omega_{ex}(r\to\infty)=0$ yields 

\begin{eqnarray}
W_m(r)  &=& -2\pi(2{\mathcal R}^2-{\mathcal R} r )\gamma_{lv} \nonumber \\
  &\:& -\frac{\pi}{12}(r^3-12{\mathcal R}^2r+16{\mathcal R}^3)\delta p\:.
  \label{eq:W(r)}
\end{eqnarray}

Both terms in (\ref{eq:W(r)}) approach zero  from below as $r\to 2{\mathcal R}^-$.  We shall refer to the first as the `area' term and the second as the `volume' term. \footnote{In principle, the value of $\ell$ appearing in Eq.\,(\ref{eq:W(r)}) depends on $r$, however, we expect this dependence to be weak and neglect it here.} Note that the fact that the grand potential per unit volume is simply the local pressure means that the form of $W_m(r)$, Eq. ~\ref{eq:W(r)}, carries over to the isobaric ensemble in which we conduct simulations.

This schematic model connects the effective potential $W(r)$ to surface tension $\gamma_{lv}$, oversaturation $\delta p$, and a characteristic solvation shell width $\ell$. It has been derived assuming the solvation shell is a true vapor layer such that $\gamma_{lv}$ and $\delta p$ take their bulk values~\cite{Coe2023}. However, in our case, an inhomogeneous layer forms, distinct from the bulk vapor phase, as shown in Figs.~\ref{fig:rho-chi-Rs10} and \ref{fig:rho-chi-Rs15}. This motivates us to consider {\em effective} values of the surface tension, and pressure difference $\delta p$ when comparing Eq.\,(\ref{eq:W(r)}) to simulations, which we denote by $\tilde\gamma, \delta \tilde p$ respectively. Together with the value of $\ell$, these effective values will be determined as independent fitting parameters.

\section{Scaling properties of the solvation layer width $\ell$}

Variations in $\ell$ modify $\mathcal{R} = R_s + \ell$ and its powers in the surface and volume terms of Eq. (\ref{eq:W(r)}), making $\ell$ central to the range and depth of the effective potential $W(r)$. Its dependence on pressure, temperature, solute size, and hydrophobicity is therefore key.

A mean-field effective interface potential model~\cite{Coe2023} links $\ell$ to these parameters via the excess free energy per unit area, $\omega_{ex}(\ell)$, of a vaporlike solvation layer. 
Typically, $\omega_{ex}(\ell)$ consists of three $\ell$-dependent contributions to the interaction between the solute and the outer surface of the solvation shell:

\begin{enumerate}
\item[(i)] a repulsive ``missing neighbor'' term from the energetic cost of liquid near a purely repulsive solute, 

\item[(ii)] an attractive term from solute-water dispersion forces favoring liquid contact,

\item[(iii)] an additional attractive term from the free energy cost of sustaining a metastable vapor layer and the Laplace pressure due to interfacial curvature.

\end{enumerate}

The equilibrium shell width $\ell$ minimizes the total free energy from these competing terms. The form of $\omega_{ex}(\ell)$ depends on the range and shape of solvent-solvent and solute-solvent interactions~\cite{Coe2023}, with implications for the scaling of $\ell$ (Sec.~\ref{sec:scaling}). For further details and examples of binding potentials, see~\cite{Coe2023}.

\subsubsection{Purely repulsive solute-water interactions}

For an isolated extended solute that is purely repulsive to the solvent (i.e., maximally hydrophobic), and where solvent-solvent interactions are short-ranged,  binding potential arguments \cite{Coe2023} predict that the width $\ell$ of a vaporlike solvation layer scales according to 

\begin{equation}
\ell =\xi_b\ln(\beta\sigma^3a(T)/\xi_b) -\xi_b \ln \left(\beta \sigma^3 \delta p^\prime\right)\:.
\label{eq:SRLRelleq}
\end{equation}
Here $\xi_b$ is the bulk vapor phase correlation length, $a(T)$ is a positive,  weakly temperature-dependent background term, $\sigma$ is the water molecular diameter, and $\beta=1/k_BT$.  $\delta p^\prime$ is a scaling field that takes the form of an effective oversaturation that accounts for curvature effects:

\begin{equation}
\delta p^\prime  = p_l-p_v + \frac{2 \gamma_{lv}}{R_s}\:,
\label{eq:deltapt}
\end{equation}
with $p_l$ the pressure of bulk liquid water and $p_v$ the vapor pressure at the chemical potential of the liquid. Since the bulk liquid is oversaturated, the vapor phase is metastable. The second term on the right-hand side of Eq.\,(\ref{eq:SRLRelleq}) is a Laplace pressure which captures the effect of curvature and acts as an oversaturation. Eq.\,(\ref{eq:SRLRelleq}) encapsulates the physics of the critical drying transition: as $R_s \to \infty$ and the bulk pressure approaches the solvent's vapor-liquid coexistence pressure (i.e., $p_l \to p_v$), then from Eq.\,(\ref{eq:deltapt}), $\delta p^\prime\to 0$, and the film thickness $\ell$ diverges logarithmically.

Eqs.\,(\ref{eq:SRLRelleq},\ref{eq:deltapt}) thus relate changes in $\ell$ to changes in temperature, pressure, and solute radius. Specifically: 

\begin{enumerate}
    \item  Rising $T$ at constant $p$ leads to an increase in $\ell$ because $\beta$,$\gamma_{lv}$ and the oversaturation $\delta p$ all decrease.
    \item  Increasing bulk pressure $p_l$ leads to a reduction in $\ell$.
    \item Increasing $R_s$ leads to an increase in $\ell$.
\end{enumerate}  
We test these predictions in the simulations reported in Sec.~\ref{sec:simulation}.

\subsubsection{Weakly attractive solute-water interactions}

Real apolar solutes attract water weakly via dispersion forces~\cite{Rein_ten_Wolde:2002}, with the solute–water well depth $\varepsilon_{sw}$ serving as an inverse measure of hydrophobicity. Binding potential studies show that increasing $\varepsilon_{sw}$ modifies the solvation layer width and introduces an additional positive $T$- and $\ell$-dependent contribution to $\delta p^\prime$, denoted $g(\varepsilon_{sw},\ell)$~\cite{CoeEvansWildingPRL2022,Coe2023}:

\begin{equation}
\delta p^{\prime\prime}= p_l-p_v + \frac{2\gamma_{lv}}{R_s} + g(\varepsilon_{sw},\ell)\:.
\label{eq:tildepattract}
\end{equation}

Here $g(\varepsilon_{sw},\ell)$ acts as an effective inward `attraction pressure’ on the solvation shell–bulk interface, proportional to $\varepsilon_{sw}$~\cite{Coe2023,CoeThesis}. As $\ell$ grows and bulk liquid recedes from the solute, this pressure decreases, making $g(\varepsilon_{sw},\ell)$ a monotonically decreasing function of $\ell$.

The precise form of the $\ell$ dependence of $g(\varepsilon_{sw},\ell)$ rests on the asymptotic character of the solute-solvent and solvent-solvent interactions, eg. for long-ranged solute-water dispersion interactions, one finds  $g(\varepsilon_{sw},\ell)=b(T)\varepsilon_{sw}\ell^{-3}$ with $b(T)=b_0(\varrho_l-\varrho_v)$, a weakly decreasing function of $T$~\cite{Coe2023}. For truncated dispersion interactions, or an exponentially decaying solute-water interaction potential, $g(\varepsilon_{sw},\ell)$ is expected to decay exponentially with $\ell$. However, it will be sufficient for our immediate purposes that $g(\varepsilon_{sw},\ell)$ is positive and a decreasing function of $\ell$ and of $T$.   

With this addition, Eq.\,(\ref{eq:SRLRelleq}) is modified to read
  \begin{equation}
     \ell=\xi_b\ln(\beta\sigma^3 a(T)/\xi_b) -\xi_b\ln(\beta\sigma^3\delta p^{\prime\prime}),
     \label{eq:SRLRellattract}
 \end{equation}
where, as in the case of purely repulsive solute-water interactions, the second term on the right-hand side of Eq.\,(\ref{eq:SRLRellattract}) captures the singular behaviour of $\ell$. Eqs.\,(\ref{eq:tildepattract},\ref{eq:SRLRellattract}) show that as far as the value of $\ell$ is concerned, the role of solute-water attraction is equivalent to increasing the oversaturation in a (maximally hydrophobic) system having purely repulsive solute-water interactions.

As we will demonstrate, despite the absence of a distinct vapor-like solvation layer structure for solutes with a radius of $R_s \approx 10$\,{\AA}, simulations validate the predicted trends for $\ell$ for both purely repulsive and weakly attractive solute-water interactions. By integrating the above scaling theory for $\ell$ with our morphometric model, we thus establish a robust framework for understanding the key factors influencing the detailed form of the hydrophobic attraction.

\section{Simulation results for the effective potential:  $W(r)$}
\label{sec:simulation}

Isothermal-isobaric MD simulations (see Appendix~\ref{app:mD}) have been conducted to measure $W(r)$ for pairs of solute particles immersed in the liquid phase of mW water~\cite{Molinero:2009aa,CoeEvansWildingJCP2022} at various pressures and temperatures. We have considered the case of purely repulsive solutes as well as the impact of introducing an isotropic attractive solute-water potential, $u_{sw}(r_{sw})$.

Direct simulations of pairs of spherical solutes under ambient conditions can result in almost irreversible aggregation, so we utilized long metadynamics simulations, a well-established method for measuring $W(r)$~\cite{Barducci:2008}. In the following, we will describe our simulation results and compare them with the model predictions for $W_m(r)$, and $\ell$ given by Eq.\,(\ref{eq:W(r)}) and Eqs.\,(\ref{eq:tildepattract},\ref{eq:SRLRellattract}). 

\subsection{Purely repulsive solutes}

\label{sec:purelyrepulsive}

We model an idealized, maximally hydrophobic solute using a purely repulsive Morse potential for both solute-solute and solute-water interactions as detailed in Appendix~\ref{app:interpot}. Fig.~\ref{fig:pmfsT300}(a) and (b) show the effective potential   $W(r)$ for $R_s=10\,${\AA} and $R_s=15\,${\AA} respectively, at temperature $T=300$K for several pressures in the range $0\le p\le 1500\,$atm. Differences between $p=0$ and ambient pressure of $p=1\,$atm are unresolvable on the scale of the pressure effects on $W(r)$ that we observe, thus $p=0$ provides a good description of both coexistence and ambient conditions. To compare between the two values of $R_s$, we plot the potentials versus $r^\prime= r - 2R_s$ and scale their magnitude by $k_B T$.

For both $R_s=10\,${\AA} and $R_s=15\,${\AA} we observe the following features: (i) for lower pressures, the potentials are strongly attractive on the scale of $k_BT$- a well-known characteristic of interactions between solvophobes of various shapes and sizes \cite{LumChandlerWeeks1999,Lam2017,Hopkins:2009,ChackoEvansArcher2017};  (ii) when $p=0\:$atm, the potential is approximately linear over much of the range of $r$ beyond the minimum, implying a separation-independent attractive force; (iii) On increasing $p$ (and thus the oversaturation), the potential becomes appreciably more rounded over a given range of $r$, while the attractive range and depth decrease substantially; (iv) At large $r$, $W(r)$ exhibits weak oscillatory behavior which becomes more pronounced at high pressures (see also \cite{Ghosh:2001aa,Sarma:2012}). Similar oscillations have previously been studied in the context of entropic depletion interactions in highly size-asymmetrical hard sphere mixtures~\cite{Roth2000a}, where it was shown that the radial distribution function of the pure solvent sets the oscillation period.

The main differences between the potentials for the two solute radii are that the attractive range is substantially greater for $ R_s = 15 \, \text{\AA} $ compared to $ R_s = 10 \, \text{\AA} $. Additionally, for a given pressure, the depth of the potential well more than doubles when $ R_s $ increases by $50\%$. 

It is instructive to compare $W(r|R_s=15\rm{\AA},p)$ with the corresponding local density profiles $\varrho(r|R_s=15\rm{\AA},p)$ shown in Fig.~\ref{fig:rho-chi-Rs15}(a). Somewhat surprisingly, an effective solute-solute attraction occurs even for pressures where the local water density near the solute surface exceeds $\varrho_b$, and the total adsorption $\Gamma(p)$ (inset of Fig.~\ref{fig:rho-chi-Rs15}a) is positive. This implies that a solvation shell of raised free energy can exist without an obvious signature of density depletion in the density profile. This finding can be understood in the context of the solvation layer structure discussed earlier, where, for our values of $R_s$, the solvation layer comprises a population of vapor-like bubbles within the liquid phase~\cite{CoeEvansWildingPRL2022,Coe2023}. The presence of these vapor bubbles increases the free energy relative to the bulk. However, if their population is small, $\varrho(r)$ will not significantly deviate from the form it takes for a pure liquid in the presence of a spherical substrate. In this case, the liquid displays layering features whereby -- close to the substrate -- $\varrho(r)$ can exceed $\rho_b$. Of course, a clear density depletion compared to $\rho_b$, evolving into a vapor layer, does emerge close to vapor-liquid coexistence for a purely repulsive solute as $R_s\to\infty$, i.e., for sufficiently small $\delta p^\prime$ \cite{Coe2023}. 

\begin{figure}
    \centering
    \includegraphics{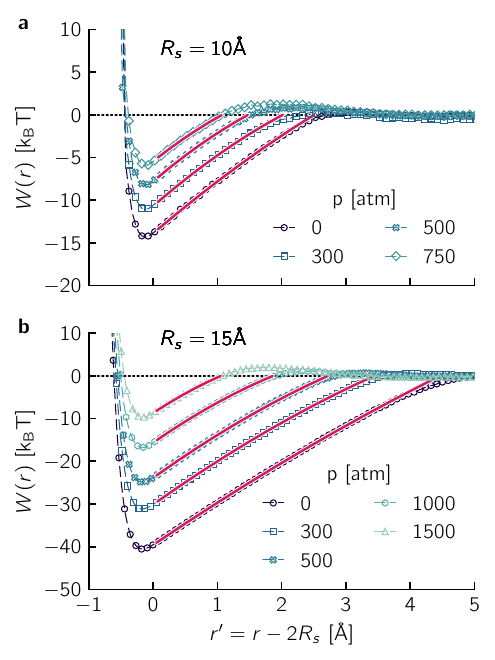}
    \caption{ Effective potential $W(r)$ (symbols) in units of $k_BT$ between two purely repulsive solutes of radii {\bf (a)} $R_s=10\,${\AA}. and {\bf (b)} $R_s=15\,${\AA} at a selection of pressures $p$. The temperature is $T=300\,$K. Fits to the potential beyond the minimum based on Eq.\,(\ref{eq:W(r)}) are shown as lines; the fit parameters $\delta \tilde p$, $\tilde\gamma$, and $\ell$ are listed in Tab.~\ref{tab:fitparamshard}.}
    \label{fig:pmfsT300}
\end{figure}

Let us assess how the qualitative features of the results for $W(r)$, Fig.~\ref{fig:pmfsT300}, accord with our model. 
\begin{itemize}
    \item[(i)] Near bulk vapor-liquid coexistence $\delta p$ is small, and the cubic volume term in Eq.\,(\ref{eq:W(r)}) becomes neglible.  This implies that $W(r)$ is principally prescribed by the area term in (\ref{eq:W(r)}) i.e., it is {\em linear} in $r$ with gradient $2\pi(R_s+\ell)\gamma_{lv}$. Our simulation results for $W(r)$ at $p=0$ are indeed approximately linear.

\item[(ii)] The attractive range of $W(r)$ is set by $\ell$, and according to Eqs.\,(\ref{eq:deltapt}) and (\ref{eq:SRLRelleq}), this grows with the solute radius $R_s$ at a given pressure, as confirmed by comparing the results of Fig.~\ref{fig:pmfsT300}(a) and (b).

\item[(iii)] The increase in $\ell$ with $R_s$ at a given pressure renders the effective potential significantly more attractive across its range due to the dependence of Eq.\,(\ref{eq:W(r)}) on ${\mathcal R}=R_s+\ell$. We elaborate on this point below.

\item[(iv)] A pressure increase raises $\delta p$, which amplifies the relative contribution of the cubic volume term in Eq.\,(\ref{eq:W(r)}). This results in more rounded (less linear) potentials over a given range of $r$. However, increasing $p$ reduces $\ell$ via Eq.\,(\ref{eq:SRLRelleq}). This leads to a substantial reduction in the magnitudes of both the volume and area terms and consequently, the depth of $W(r)$ decreases, consistent with the simulation results.

\end{itemize}

We now evaluate the accuracy of Eq.\,(\ref{eq:W(r)}) in quantitatively describing the observed forms of $W(r)$. To achieve this, we selected the appropriate value of $R_s$ and optimized the effective model parameters $\delta \tilde p$, $\tilde{\gamma}$, and $\ell$ to best match the simulation results. These fits focus on the region beyond the potential minimum and extend until just before the intersection with the $W(r) = 0$ axis, as outlined below. The resulting curves, shown in Fig.~\ref{fig:pmfsT300}, provide an excellent description of the measured potential across all pressure values. The corresponding parameters are listed in Table~\ref{tab:fitparamshard}.

\begin{table}[t!]
\centering
\begin{tabular}{lr|rrr}
$R_s$\:(\AA) & $p$ (atm) & $\delta \tilde p$ (atm) &
$\tilde{\gamma} \:(N/m)$& $\ell$\:(\AA) \\
\midrule
10 & 0.0  & 889(200) & 0.027(2) & 1.25(1) \\
\: & 300  & 1576(200) & 0.024(2) & 1.01(1) \\
\: & 500  & 2022(200) & 0.025(1) & 0.74(1) \\
\: & 750  & 2205(300) & 0.026(1)  & 0.54(2)\\\hline
15 & 0    & 488 (100) & 0.029(3) & 2.23(1) \\
\: & 300  & 935 (100) &  0.027(3) & 1.72(1) \\
\: & 500  & 1235 (200) & 0.027(2) & 1.36(1) \\
\: & 1000 & 1933(200) &  0.026(3) & 0.93(4) \\
\: & 1500 & 2776 (300) & 0.028(3) & 0.53(4) \\
\bottomrule
\end{tabular}
\caption{Least squares fit parameters for $\tilde\gamma$, $\delta \tilde p$, and $\ell$ based on the model in Eq.\,(\ref{eq:W(r)}), for purely repulsive solutes with radii $R_s=10\,${\AA} and $R_s=15\,${\AA} at the specified pressures. Note that $\ell$ is very small -- less than a water diameter $\sigma_{mw}=2.392${\AA}, see Appendix~\ref{app:mW}.}
\label{tab:fitparamshard}
\end{table}

Despite $R_s$ being the sole constrained parameter, the values of $\tilde\gamma$ and $\delta \tilde p$ are consistent in magnitude with the mW value of the bulk vapor-liquid surface tension $\gamma_{lv}=0.066\,$N/m at $T=300\,$K \cite{CoeEvansWildingJCP2022} and the applied pressure $p$, as indicated in the legend of Fig.~\ref{fig:pmfsT300}. This provides strong support for the assumptions underlying Eq.\,(\ref{eq:W(r)}). Further indications that validate the model and show that the fitted values are physically meaningful include (i)  $\delta \tilde p$ follows the increase in $p$; (ii) $\tilde\gamma$ is approximately $40\%$ of $\gamma_{lv}$, which is reasonable given that the solvation layer is not vapor-like; and (iii)  $\ell$ accurately reproduces the potential range of approximately $2\ell$, as confirmed by examining Fig.~\ref{fig:pmfsT300}. 

At lower pressures, close to coexistence, the fits somewhat underestimate $W(r)$ at large $r$, which we attribute to enhanced (near-critical) density fluctuations within the solvation shell causing $\ell$ to fluctuate significantly from its mean value. This feature, which is not included in our model, means that a pair of solutes will start to interact on a length scale somewhat larger than the average $\ell$ that results from a fit of $W(r)$ with Eq.\,(\ref{eq:W(r)}). For this reason, we do not attempt to fit right up to the $W(r)=0$ axis but over the range $2R_s<r\lesssim 2R_s+1.9 \ell$

\begin{figure}[t]
    \includegraphics{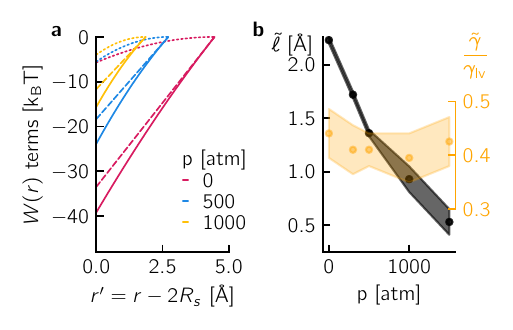}
\caption{{\bf (a)} Contributions of the volume (dotted line) and area (dashed line) terms in Eq.\,(\ref{eq:W(r)}) to $W_m(r)$ (continuous line) for purely repulsive solutes with $R_s=15\,${\AA}, $T=300\,$K, corresponding to the fits of Fig.~\ref{fig:pmfsT300}(b) with parameters listed in Tab.~\ref{tab:fitparamshard}. Results are presented for the fits to simulation data for $p=0$ atm (red), $p=500$ (blue), and $p=1000$ (orange) and are discussed in the text. {\bf (b)} Pressure dependence of the fit parameters for $R_s=15\,${\AA}, $T=300\,$K of $\tilde\gamma/\gamma_{lv}$ (orange), and $\ell$ (black), see also Tab.~\ref{tab:fitparamshard}. Shaded areas are uncertainties.}
\label{fig:modelcontribs}
\end{figure}

Figure~\ref{fig:modelcontribs}(a) depicts the contributions of the individual volume and area terms to the model fits for $W(r)$ (Eq.\,\ref{eq:W(r)}), focusing on $R_s=15\,${\AA} at pressures of $p=0$, $500$, and $1000$~atm. The figure demonstrates that the linear area term is the dominant factor, which partly accounts for the larger uncertainties in the estimates of $\delta \tilde p$ compared to those of $\tilde\gamma$ in Tab.~\ref{tab:fitparamshard}. However, as $\delta \tilde p$ increases, the relative importance of the volume term grows significantly, resulting in a more pronounced rounding of $W(r)$ over a given range of $r$.

While the area term dominates $W(r)$ for purely repulsive solutes, our fits for $\tilde\gamma(p)$ and $\ell(p)$ listed in Tab.~\ref{tab:fitparamshard} and plotted in Fig.~\ref{fig:modelcontribs}(b) indicate that $\tilde\gamma$ remains approximately constant as $p$ is varied. By contrast, $\ell$ increases substantially as $p$ is lowered. This sensitivity of $\ell$ to deviations in $p$ from vapor-liquid coexistence accords with Eq.\,(\ref{eq:SRLRelleq}) and stems from critical drying. The observed strong variation in the attractive strength of $W(r)$ with changing $p$ (as shown in Fig.~\ref{fig:pmfsT300}) is thus primarily driven by the changes in $\ell$. Indeed, the dominant contribution of the area term to $W(r)$ at contact ($r=2R_s$) scales as $\tilde \gamma(R_s\ell+\ell^2)$ and is thus very sensitive to changes in $\ell$. Of course, this underscores the importance of understanding what determines the magnitude of $\ell$.

Finally, in this subsection, we note that a smooth, nearly linear $W(r)$ can arise from overlapping solvation shells even when $\rho(r)$ is highly structured, with oscillations extending well beyond fitted $\ell$ (cf. Figs. \ref{fig:pmfsT300}(a,b) and \ref{fig:rho-chi-Rs10}(a), \ref{fig:rho-chi-Rs15}(a)). Thus, one should not generally expect a clear signature of $\ell$ in $\rho(r)$. In the limit of a vapor-like drying film, $\ell$ is identifiable in $\rho(r)$ as the radial extent of the low-density region~\cite{Coe2023}, but for the radii studied here no such unambiguous feature appears: the solvation layer is a fluctuating mix of liquid- and vapor-like regions, and packing-induced oscillations from the liquid-like portions obscure $\ell$.

The smooth $W(r)$ indicates that packing effects do not necessarily elevate the solvation shell free energy above bulk water. Oscillations in $\rho(r)$ result from measuring with the solute at the origin, which produces structure whenever the shell is not vapor-like, analogous to $g(r)$ in a bulk liquid where oscillations occur despite uniform free-energy density and no net attraction. Nonetheless, a signature of $\ell$ in $\rho(r)$ that evolves toward the width of a vapor layer near critical drying can be detected, as illustrated in Fig. 3 of~\cite{WildingTurciPRR2025}.

\subsection{Weakly attractive solute-water interactions}

We have performed simulations of solutes with attractive solute-water interactions employing a Morse potential truncated at a large radius (see Appendix~\ref{app:interpot} for details). We find that our solutes of radius $R_s=10${\AA} exhibit hydrophobic behavior (defined as the existence of solute-solute attraction) for $0 < \varepsilon_{sw} \lesssim 0.7$ kcal/mol and we have considered attractive strengths of $\varepsilon_{sw}=0.2,0.4,0.6$ kcal/mol. For reference,  at $T = 300$K, $1$ kcal/mol$=1.677k_B T$. 

\begin{figure}
    \centering
    \includegraphics{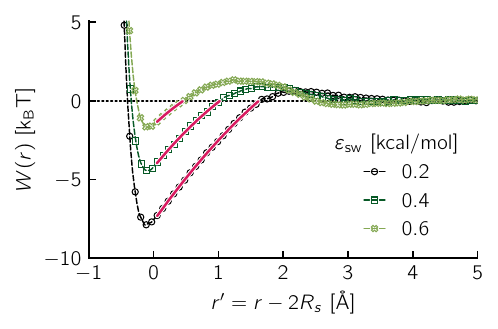}
    \caption{Effective potential $\beta W(r)$  for $R_s=10\,${\AA} at $p=0$ atm, $T=300\,$K for solute-water well depths $\varepsilon_{sw}=0.2,0.4,0.6$ kcal/mol. Fits based on Eq.\,(\ref{eq:W(r)}) to the attractive part of the potential beyond the solute diameter are shown as solid lines; fit parameters are plotted in Fig.~\ref{fig:l-gamma-dp}. Note that at $T=300$K $1$ kcal/mol=1.677$k_BT$  and $\gamma_{lv}=0.066\,$N/m for mW \cite{CoeEvansWildingJCP2022}.}
    \label{fig:pmf_D0_at_T300K}
\end{figure}

The effect of varying $\varepsilon_{sw}$ on $W(r)$, is shown in Fig.~\ref{fig:pmf_D0_at_T300K} for $p=0$~atm (which, recalling the discussion of sec.~\ref{sec:purelyrepulsive}, corresponds essentially to vapor-liquid coexistence in our model). Qualitatively, as $\varepsilon_{sw}$ increases, the range of $W(r)$ diminishes strongly, its depth becomes much shallower and the potential form becomes increasingly rounded compared to the purely repulsive solute at $p=0$, shown in Fig.~\ref{fig:pmfsT300}(a). There is also a significant decrease in the potential gradient.

To investigate the effects of $\varepsilon_{sw}$ on the form of $W(r)$ within our model, we fitted the data from Fig.~\ref{fig:pmf_D0_at_T300K} using Eq.\,(\ref{eq:W(r)}), with parameters $\delta \tilde p$, $\tilde \ell$, and $\tilde\gamma$. The fits are shown as solid lines in Fig.~\ref{fig:pmf_D0_at_T300K} and the parameters are plotted in Fig.~\ref{fig:l-gamma-dp} and listed in Tab.~\ref{tab:fitparamsattract}. We find that $\delta \tilde p$ increases rapidly with $\varepsilon_{sw}$ resulting in an increased relative contribution from the volume term in Eq.\,(\ref{eq:W(r)}). The solvation shell width $\ell$ decreases in line with the predictions of Eqs.\,(\ref{eq:tildepattract},\ref{eq:SRLRellattract}) given the presence of the attraction pressure term which is proportional to $\varepsilon_{sw}$.  We further find that $\tilde\gamma$ decreases significantly with increasing  $\varepsilon_{sw}$ reflecting the observed change in gradient of $W(r)$, and suggesting that solute-water attraction increases the effective density in the solvation shell, thereby reducing the density difference to bulk water which controls $\tilde\gamma$. 

A noticeable feature of Fig.~\ref{fig:pmf_D0_at_T300K} is a repulsive hump in $W(r)$, which grows with $\varepsilon_{sw}$ and shows signs of oscillatory behavior at large $r$. This has been referred to in the literature as a ``desolvation barrier''~\cite{Ghosh:2023}. The phenomenon appears to be a packing effect, presumably related to that which occurs in highly size-asymmetrical hard-sphere mixtures~\cite{Roth2000a} but amplified by the solute-water attraction. The effect of attractions between different hard sphere species has previously been investigated \cite{Louis:2002fk},  where the repulsive barrier was termed `accumulation repulsion'.  This repulsive effect is not included in our theory for the hydrophobicity-induced attraction between solutes.  However, the fits in Fig.~\ref{fig:pmf_D0_at_T300K} are concerned with the part of the potential that is net attractive and should not be greatly affected by this omission.

\subsection{Solute-solute attraction increases with temperature}
\label{sec:inverseT}

\begin{figure}
    \centering
    \includegraphics{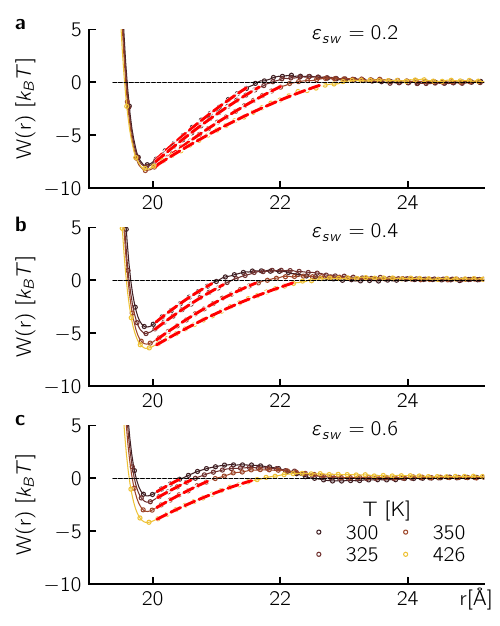}
    \caption{Effective potential $\beta W(r)$ for $R_s=10\,${\AA} at $p=0$~atm and the temperatures listed in the legend. The solute-water attractive well depth is {\bf (a)} $\varepsilon_{sw}=0.2$ kcal/mol, {\bf (b)} $\varepsilon_{sw}=0.4$ kcal/mol. {\bf (c)} $\varepsilon_{sw}=0.6$ kcal/mol. Fits based on Eq.\,(\ref{eq:W(r)}) to the attractive part of the potential beyond the solute diameter are shown as dashed lines; fit parameters appear in Fig.~\ref{fig:l-gamma-dp}.}
    \label{fig:pmf_T_variousD}
\end{figure}

Fig.~\ref{fig:pmf_T_variousD} shows $W(r)$ at $p=0$ atm for $R_s=10\,${\AA} at four temperatures: $T=300$K, $325$K, $350$K, and $426$K, and three values of $\varepsilon_{sw}=0.2, 0.4, 0.6$ kcal/mol. For all three values of $\varepsilon_{sw}$, a pronounced $T$-dependence is observed, with $W(r)$ becoming more attractive and considerably longer-ranged as $T$ rises. This is a demonstration on microscopic scales of the celebrated `inverse' temperature dependence of the hydrophobic effect.

The observed $T$-dependence is explained by our model, specifically Eqs.\,(\ref{eq:W(r)},\ref{eq:tildepattract}-\ref{eq:SRLRellattract}). Consider first the effective oversaturation $\delta p^{\prime\prime}$ that controls the value of $\ell$ via Eq.\,(\ref{eq:tildepattract}). As $T$ increases, the argument of the logarithm decreases for several reasons: (i) it contains an overall factor of $\beta=(k_BT)^{-1}$; (ii) $p_l-p_v$ decreases because it vanishes at coexistence and heating ambient water reduces the deviation from the boiling point; (iii) $\tilde\gamma$ decreases because it inherits the known $T$-dependence of the bulk surface tension $\gamma_{lv}$ which is a decreasing function of temperature~\cite{CoeEvansWildingJCP2022}; (iv) $g(\varepsilon_{sw},\ell)$ decreases with rising $T$ and additionally is a strongly decreasing function of $\ell$ \cite{Coe2023}. 

Together, these $T$-dependences result, via Eq.\,(\ref{eq:SRLRellattract}), in an increase in the solvation shell width $\ell$, and thus the attractive range of $W(r)$. Importantly,  this growth in $\ell$ is amplified by the logarithmic form of Eq.\,~\ref{eq:SRLRellattract}, which varies rapidly for small values of the argument $\delta p^{\prime\prime}$. This amplification stems from the critical divergence of $\ell$ for $\delta p^{\prime\prime}=0$.

\begin{table}[t]
\centering
\begin{tabular}{lr|rrr}
 $\varepsilon_{sw}$ (kcal/mol) & $T$ (K) & $\delta \tilde p$ (atm) & $\tilde\gamma \:(N/m)$ & $\ell$\:(\AA) \\
\midrule
0.2 & 300 & 1713(150)& 0.022(1) & 0.81(2) \\
\: & 325 & 1581(100) & 0.021(1) & 0.91(2) \\
\: & 350 & 1485(200) & 0.018(2) &1.09(3) \\
\: & 375 & 1185(150) & 0.019(2) & 1.11(3) \\
\: & 400 & 1209(150) & 0.017(2) & 1.31(3) \\
\: & 426 & 1181(200) &  0.016(1) & 1.37(2)\\\hline
0.4 & 300 & 2338(200) & 0.020(1) & 0.51(2) \\
\: & 325 &  1885(150) & 0.020(1) & 0.62(3) \\
\: & 350 &  1727(150) & 0.017(2) &0.86(2) \\
\: & 375 & 1339(150) & 0.018(2) & 0.91(3) \\
\: & 400 & 1390(150) & 0.017(2) & 1.03(3) \\
\: & 426 & 1378(200) &  0.015(1) & 1.18(3)\\\hline
0.6 & 300 & 3163(400) &  0.018(1) & 0.23(3)\\
\:  & 325 & 2022(300) &  0.018(1) & 0.33(2)\\
\:  & 350 & 1969(200) &  0.017(1) & 0.49(2)\\
\: & 375 & 1903(200) & 0.016(2) & 0.61(3) \\
\: & 426 & 1747(200) &  0.014(1) & 0.85(1)\\
\bottomrule
\end{tabular}
\caption{Least squares fits for $\tilde\gamma,\delta \tilde p$ and $\ell$ based on Eq.\,(\ref{eq:W(r)}) for solutes of radii $R_s=10\,${\AA} at $p=0$ atm and indicated values of $T$ and solute-water attractive well-depth $\varepsilon_{sw}$. }
\label{tab:fitparamsattract}
\end{table}

\begin{figure}[tbh]
\centering
\includegraphics{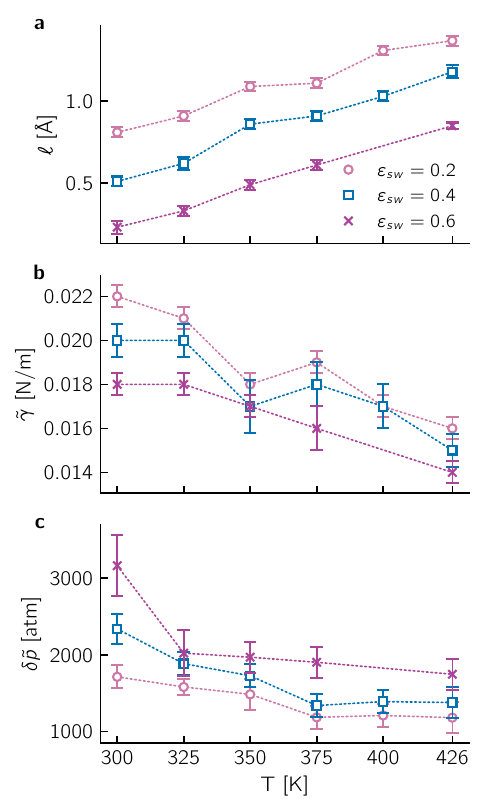}
\caption{Plots of the fit parameters $\ell$, $\tilde\gamma$, and $\delta \tilde p$ (as listed in Tab.~\ref{tab:fitparamsattract}) as determined from least squares fits to the measured forms of $W(r)$, using the model given in Eq.\,(\ref{eq:W(r)}). These parameters pertain to solutes with radii $R_s = 10$,{\AA} at ambient pressure ($p = 0$ atm), for the specified temperatures $T$ and solute-water attraction strengths $\varepsilon_{sw}$ (in kcal/mol). 
 }
\label{fig:l-gamma-dp}
\end{figure}

Fits of the form Eq.\,(\ref{eq:W(r)}) to the data of Fig.~\ref{fig:pmf_T_variousD} corroborate these trends in $\ell,\tilde{\gamma}$ and $\delta \tilde p$ with rising $T$. The temperature dependence of the fit parameters is plotted in Fig.~\ref {fig:l-gamma-dp} and listed in Tab.~\ref{tab:fitparamsattract}. We note that the magnitude of the measured decrease in $\tilde{\gamma}$ of $\approx 25\%$ in the studied range of $T$ is comparable to the decrease by circa $18\%$ of the bulk surface tension $\gamma_{lv}$ measured for mW~\cite{CoeEvansWildingJCP2022}.  The decreases in $\tilde{\gamma}$ and $\delta \tilde p$ with rising $T$ (together with that in $g(\varepsilon_w,\ell)$ demonstrated previously~\cite{Coe2023}) drive a much stronger increase in $\ell$, of $70\%$ for $\varepsilon_{sw}=0.2$, $130\%$ for $\varepsilon_{sw}=0.4$, and $270\%$ for $\varepsilon_{sw}=0.6$, which in turn acts to render $W(r)$ more attractive over the entire range. Although the absolute value of $\ell$ remains modest relative to the size of a solvent molecule (as discussed in Sec.~\ref{sec:solvation}), in percentage terms it increases sharply with $T$, in line with the prediction of scaling theory as we show below. 

Note that the decrease in $\tilde\gamma$ as temperature rises primarily affects the slope of the dominant linear (area) term in $W(r)$, making the attractive tail of $W(r)$ flatter. However, the overall attraction increases with rising temperature because the increase in $\ell$ outweighs the flattening effect. Specifically, consider the minimum value of the model potential $W_m(r)$, given by

\begin{equation}
    W_m(2R_s) = -2\pi\delta p( R_s\ell^2 + 2\ell^3/3 ) - 4\pi\gamma_{lv}( R_s\ell + \ell^2 )\:.
    \label{eq:Wmin}
\end{equation}
Substituting the values of $\delta \tilde p(T)$, $\tilde\gamma(T)$, and $\ell(T)$ from Tab.~\ref{tab:fitparamsattract} into Eq.\,(\ref{eq:Wmin}) confirms that the two terms in parentheses on the right-hand side increase more rapidly with increasing temperature than their respective prefactors decrease. This stems essentially from the rapid growth of $\ell$ with $T$, coupled with the geometry of the problem, which leads to terms involving first, second and third powers of $\ell$ in Eq.\,(\ref{eq:Wmin}).  The overall effect is that $\beta W(r)$ becomes more attractive as the temperature rises.

\begin{figure}[t]
\centering
\includegraphics{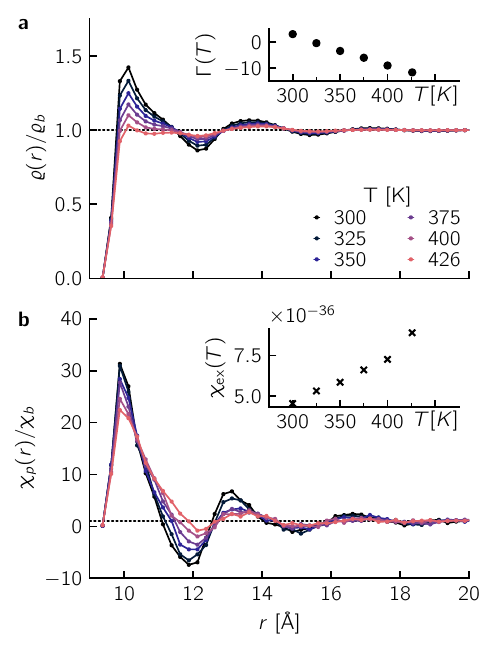}
\caption{Temperature dependence of local density and compressibility profiles for mW water in contact with a solute of radius $R_s=10${\AA} at $p=0$. The attractive solute-water potential well depth is $\varepsilon_{sw}=0.2$ kcal/mol. (a) $\varrho(r)/\varrho_b$ (b) $\chi_p(r)/\chi_p(b)$. Insets show the temperature dependence of  (a) the dimensionless total surface adsorption $\Gamma$ and (b) surface excess compressibility $\chi_{ex}$ in units of  $J^{-1}m^3$.}
\label{fig:profs0.2}
\end{figure}

\begin{figure}[h]
\centering
\includegraphics{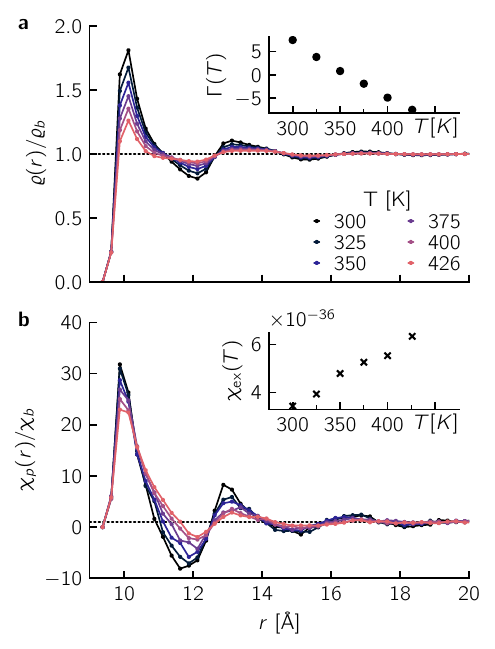}
\caption{As for Fig.~\protect\ref{fig:profs0.2} but for $\varepsilon_{sw}=0.4$ kcal/mol.  }
\label{fig:profs0.4}
\end{figure}

The increase in hydrophobic attraction with rising temperature correlates with indicators of the effective degree of hydrophobicity for an isolated sphere, as measured by $\varrho(r)$ and $\chi_p(r)$.  Figures~\ref{fig:profs0.2} and \ref{fig:profs0.4} illustrate the temperature dependence of $\varrho(r)$ and $\chi_p(r)$ at $p = 0$ atm for $R_s = 10${\AA}, with $\varepsilon_{sw} = 0.2$ kcal/mol and $\varepsilon_{sw} = 0.4$ kcal/mol respectively. These conditions correspond to the $W(r)$ measurements in Fig.~\ref{fig:pmf_T_variousD}(a) and (b). The measurements show that the local density near the solute surface decreases sharply as temperature increases, resulting in a reduction in the total surface adsorption, $\Gamma = 4\pi\int r^2\,dr\, (\varrho(r)-\varrho_b)$, which becomes negative. The width of the first peak in $\chi_p(r)$ broadens significantly with rising $T$ and the surface excess compressibility, $\chi_{ex} = 4\pi\int r^2 dr (\chi_p(r) - \chi_b)$, similarly increases. These trends are consistent with an increase in the effective degree of hydrophobicity of the solute as $T$ rises.  

\begin{figure}[h]
\centering
\includegraphics{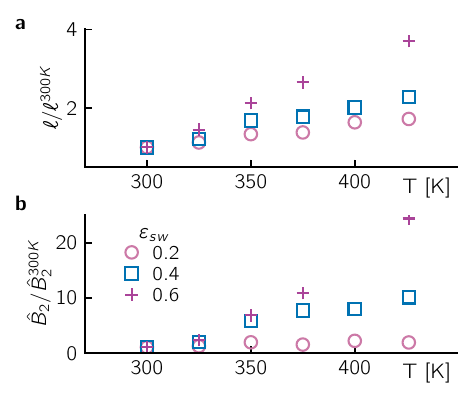}
 \caption{Temperature dependence of the relative changes in (a) the solvation shell width $\ell(T)$ and (b) the attractive component of the osmotic second virial coefficient $\hat B_2(T)$, as defined in the main text. The reference temperature is $T = 300\mathrm{K}$. Results are shown for three values of $\varepsilon_{sw}$.}
\label{fig:ellB2Tdep}
\end{figure}

An interesting feature of the $T$-dependence of $W(r)$ shown in Fig.~\ref{fig:pmf_T_variousD} is that the inverse temperature effect becomes more pronounced as $\varepsilon_{sw}$ increases, i.e., as the solute's affinity for water rises. This trend is quantified in Fig.~\ref{fig:ellB2Tdep}(a,b), which respectively show the $T$-dependence of $\ell$ and the {\em relative} change in $\ell$ for each of the three values of $\varepsilon_{sw}$ examined. Additionally, Fig.~\ref{fig:ellB2Tdep}(c) displays the relative change in $\hat B_2 \equiv -2\pi\int_{2R_s}^{r^\star} r^2 (\exp(-\beta W(r)) - 1) dr$. Here, $\hat B_2$ represents the contribution of the attractive part of $W(r)$ to the osmotic second virial coefficient, restricted to the range where $W(r)$ is negative (i.e., $2R_s < r < r^\star$, with $\beta W(r^\star) = 0$). This quantity serves as a measure of solute-solute attraction, and notably, its $T$-dependence becomes more pronounced with increasing $\varepsilon_{sw}$.

That the relative magnitude of $T$-dependent growth in the range and separation of the curves for $W(r)$ increases with $\varepsilon_{sw}$ can plausibly be explained by the functional form of the expression for $\ell$, Eqs.\,\ref{eq:tildepattract}-\ref{eq:SRLRellattract} and specifically the influence of the attraction pressure $g(\varepsilon_{sw},\ell)$ on the effective pressure $\delta{p^{\prime\prime}}$. Owing to the $\ell$ dependence of $g(\varepsilon_{sw},\ell)$, the growth in $\ell$ with increasing $T$ feeds back into $\delta{p^{\prime\prime}}$. Since the $g(\varepsilon_{sw},\ell)$ term is proportional to $\varepsilon_{sw}$, the strength of this feedback grows with $\varepsilon_{sw}$. Accordingly, the $T$-dependence of $\delta{p^{\prime\prime}}$, and hence $\ell$,  grows with $\varepsilon_{sw}$ and this is reflected in a stronger $T$-dependence of the effective potential $W(r)$.  

\subsection{Comparison with the scaling expression for $\ell$}
\label{sec:scaling}

Our estimates of the solvation shell width, $\ell$, obtained by fitting the effective interaction potential $W(r)$ across a range of $\varepsilon_{sw}$ and $T$, allow a test of the scaling predictions for $\ell$ as described by Eq.\,(\ref{eq:SRLRellattract}). Specifically, one can plot the right-hand side of Eq.\,(\ref{eq:SRLRellattract}) against $\ell$ for various combinations of $\varepsilon_{sw}$ and $T$. If the scaling holds exactly, the data should collapse onto a single linear master curve. However, pure scaling is only expected to emerge close to the critical drying limit where fluctuations are most pronounced~\cite{CoeEvansWildingPRL2022, Coe2023}. The solutes considered here, with radius $R_s = 10$~Å, do not exhibit a fully developed vapor-like solvation layer, and thus Eq.\,(\ref{eq:SRLRellattract}) is not expected to yield quantitative accuracy. Nevertheless, the scaling form should still capture the qualitative trends in how $\ell$ varies with temperature and solute–water attraction strength.

\begin{figure}[th!]
\centering
\includegraphics[width=0.85\columnwidth]{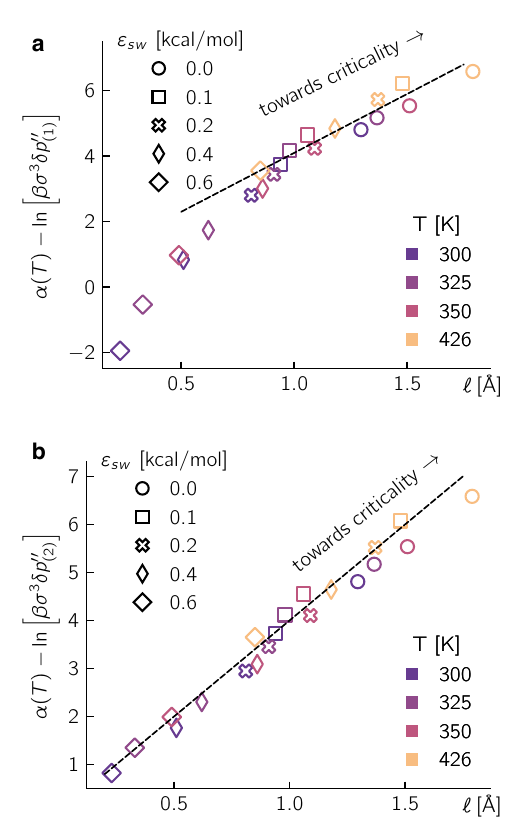}
\caption{Comparison of scaling forms.  We plot the expression Eq.\,(\ref{eq:SRLRellattract}) versus $\ell$ for various $\varepsilon_{sw}$ and $T$, and for two choices of the term $\delta p^{\prime\prime}=(p_l-p_v)+2\gamma/R_s+g(\varepsilon_{sw},\ell)$. Values of $\ell$ and $\gamma=\tilde\gamma$ are taken from Fig.~\ref{fig:l-gamma-dp} and $\sigma$ is assigned the mW value (Appendix~\ref{app:mW}).  (a)  Long-range solute-solvent scaling  $\delta p^{\prime\prime}_{(1)}$ with attraction pressure $g_{(1)}(\varepsilon_{sw},\ell)=b(T)\varepsilon_{sw}\ell^{-3}$, as derived in Ref.~\cite{Coe2023}. The quantity $b(T)=b_0(\varrho_l-\varrho_v)\sigma^6$ with  $\varrho_l-\varrho_v$ is assumed to be a constant for mW in the present temperature range \cite{CoeEvansWildingJCP2022}. (b) The same data plotted with the alternative, short-range solute-fluid ansatz $g_{(2)}(\varepsilon_{sw},\ell)=\varepsilon_{sw}\xi_b^{-3}\exp(-2\ell/\xi_b)$.  For both forms, we take   $p_l-p_v=0$ as we work close to coexistence and fit with a linear  $T$-dependent analytical background term $\alpha(T)=\xi_b\ln(\beta\sigma^3a(T)/\xi_b):=\alpha T$,  with the same coefficient $\alpha=0.01T$ empirically chosen to optimize data collapse. Dashed lines are guides to the eye.}
\label{fig:scalingplot}
\end{figure}

To construct a scaling plot, it is necessary to assume a specific form for the attractive pressure function $g(\varepsilon_{sw}, \ell)$ appearing in Eq.\,(\ref{eq:tildepattract}). Ref.~\cite{Coe2023} proposed $g(\varepsilon_{sw}, \ell) = b_0(T)\varepsilon_{sw}\ell^{-3}$, appropriate for long ranged dispersion-like interactions. Fig.~\ref{fig:scalingplot}(a) presents the corresponding scaling plot for this form. One sees that $\ell$ increases as $\varepsilon_{sw}$ decreases and $T$ increases—i.e., with increasing hydrophobicity—in agreement with the predicted scaling behaviour. Moreover, the relation becomes progressively more linear with increasing $\ell$, indicating convergence toward the pure scaling regime previously demonstrated for larger solute sizes~\cite{CoeEvansWildingPRL2022, Coe2023}.

An alternative scaling form, presumably more appropriate for the Morse potential used in the present study, which decays more rapidly than van der Waals interactions, is $g(\varepsilon_{sw}, \ell) = b_0(T)\varepsilon_{sw}\xi_b^{-3}\exp(-2\ell/\xi_b)$. Fig.~\ref{fig:scalingplot}(b) shows the corresponding scaling plot in this case. Compared to the power-law ansatz, it yields a different data collapse and provides a better match across the dataset. These results underscore that the degree of scaling collapse is sensitive to the chosen form of $g(\varepsilon_{sw}, \ell)$, emphasising the importance of considering the form of the interaction potential used when constructing a scaling ansatz.

\section{Results for an atomistic water model: SPC/E}

Similar results to those obtained with the mW water model have also been obtained using the Simple Point Charge Extended (SPC/E) model \cite{berendsen1987missing}. In this model, a water molecule is represented by distinct point charges for the hydrogen and oxygen atoms, arranged at fixed angles and including an additional polarization correction. While this correction enhances the model's accuracy, it also increases the computational cost of metadynamics simulations. As with the mW model, we introduce a weak attractive interaction, characterized by an energy scale $\varepsilon_{sw}$ between the solute and the three-point charges of the water molecule to mimic dispersion forces.

\begin{figure}[h]
    \centering
    \includegraphics
    {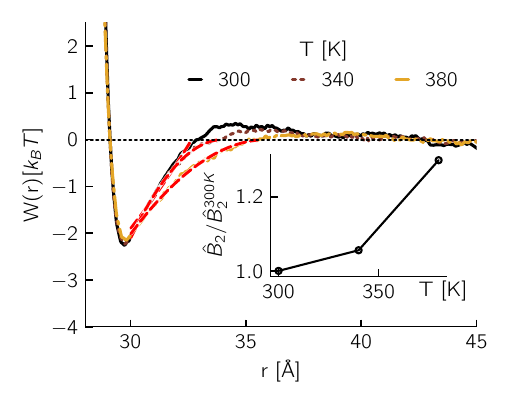}    \caption{Effective potential $W(r)$ expressed in units of $k_BT$ between a pair of solutes of radius $R=15{\rm \AA}$ in SPC/E water at three temperatures.  The solute-fluid interaction potential takes the More form described in the appendix and has strength $\varepsilon_{\rm sw}=0.4 {\rm kcal/mol}$. Dashed red lines are the theoretical model discussed in the main text. Inset: relative change in $\hat B_2$ (see main text) as a function of temperature. }
    \label{fig:spce}
\end{figure}

Figure \ref{fig:spce} presents the effective potential $W(r)/k_BT$ at three different temperatures for $\varepsilon_{\rm sw} = 0.4, {\rm kcal/mol}$, along with the corresponding fits based on the model given in Eq.\,(\ref{eq:W(r)}). Although the SPC/E and mW models differ in their physical representations, the overall form of the effective potential remains consistent across both, with temperature-dependent variations in depth and range. These differences are accurately captured by the model fit to the attractive region. The inset further illustrates the relative change in the attractive component of the second virial coefficient, highlighting the temperature dependence of the hydrophobic effect, consistent with observations from the mW model.

\section{Are hydrophobic interactions special?}
\label{sec:LJ}

Discussions of hydrophobic interactions -- and their sensitivity to temperature and pressure -- frequently focus on how changes in these fields influence the hydrogen-bond network of water. A widely held view (e.g.,~\cite{DillBromberg,israelachvili2011,Silverstein:2000aa,Widom2003,Paschek:2004,israelachvili2011,Bischofberger:2014aa,KRONBERG201614,Ghosh:2023}) posits that seemingly counterintuitive phenomena, such as the inverse temperature effect, are unique to aqueous systems and stem from changes in orientational entropy linked to the formation or disruption of hydrogen bonds with changing temperature. Yet, as mentioned in Sec.~\ref{sec:intro}, the evidence supporting this interpretation seems inconclusive, and the matter remains controversial. This raises a fundamental question: Are hydrophobic interactions truly a special case, or are they better understood as manifestations of more general solvophobic behavior? Put differently, is hydrogen bonding genuinely essential to explaining the qualitative features of hydrophobic attraction and self-assembly observed in aqueous environments?
 
To shed new light on the matter, we investigated the temperature dependence of $W(r)$ for a pair of identical spherical solvophobic solutes immersed in a simple $12$-$6$ Lennard-Jones liquid solvent. Each solute has a radius of $R_s=4\sigma$, where $\sigma$ is the Lennard-Jones solvent diameter. As for the water models, the solute-solute pair interaction is described by a purely repulsive Morse potential, while the solute-solvent interaction is described by an attractive Morse potential, with well depth $\varepsilon_{ss}=1$ (see Appendix~\ref{app:interpot}). These parameters yield a degree of solvophobicity comparable to those studied previously with the mW and SPCE water models.

\begin{figure}[t]
\centering
\includegraphics[width=0.45\textwidth]{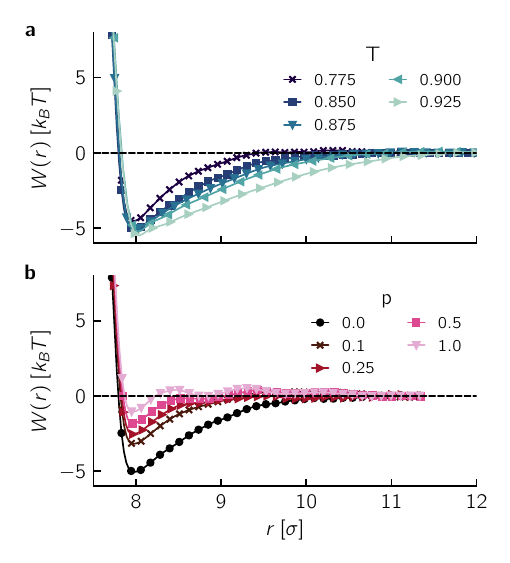}
\caption{
Measurements of $\beta W(r)$ from constant-$NpT$ well-tempered metadynamics for two solutes of radius $R_s = 4\sigma$ in a Lennard-Jones solvent of particle diameter $\sigma$, with interactions truncated at $r_c=2.5\sigma$ (bulk critical temperature $T_c=1.1876(3)$~\cite{Wilding1995}). The solute–solvent Morse potential has well depth $\varepsilon_{ss}=1.0$. (a) $\beta W(r)$ vs.~$T$ at $p=0$. (b) $\beta W(r)$ vs.~$p$ at $T=0.85$, where the coexistence pressure is $p_{coex}\approx 0.015$. Standard Lennard-Jones units are used throughout.
}
\label{fig:LJTdep}
\end{figure}

The effective pair potential, $W(r)$, was obtained as before using well-tempered metadynamics in the constant-$NpT$ ensemble, with the solvent maintained in the liquid phase across a range of temperatures and at pressures near vapor-liquid coexistence. The resulting profiles of $\beta W(r)$, reveal a strong inverse temperature dependence in the solute-solute attraction as shown in Fig.~\ref{fig:LJTdep}(a).  This trend qualitatively mirrors the behavior observed in both the mW and SPCE water models. A similar correspondence is evident in the variation of $\beta W(r)$ with pressure deviations from coexistence, as shown in Fig.~\ref{fig:LJTdep}(b).

This demonstration of the inverse temperature effect for a simple Lennard-Jones solvent supports our assertion that the temperature-induced broadening of the solvent shell, associated with a critical drying transition, is the primary mechanism for the inverse temperature effect, rather than variations in the orientational entropy of water, say. 

\section{Summary and discussion}

We investigated how solvation phenomena governed by surface phase behaviour, specifically the critical drying transition, explain solvent-mediated interactions between hydrophobic solutes and their dependence on solvent state (pressure, temperature) and solute properties (size, solvent affinity). Molecular simulations and analysis of local density and compressibility profiles around spherical solutes revealed that near vapor–liquid coexistence, a depleted low-density region forms adjacent to the solute, accompanied by amplified density fluctuations. These effects peak at coexistence, where depletion and fluctuations are maximized.

This behavior signals proximity to a critical drying transition, where a macroscopic vapor layer forms at a strongly hydrophobic planar surface. At standard temperature and pressure, water lies close to vapor–liquid coexistence, so extended hydrophobic solutes with sizes $R_s \gtrsim 10$ Å are strongly affected. The influence appears as water depletion around the solute—a precursor to the macroscopic vapor layer that emerges for very large radii—and as amplified density fluctuations. 

We quantified enhanced fluctuations via the local compressibility $\chi_p(r)$ and surface excess compressibility $\chi_{ex}(p)$, both far exceeding bulk values. Microscopically, this reflects fluctuating vapor bubbles within the solvation layer, which lie flat against the solute surface as predicted by drying theory. Correlation lengths are anisotropic, $\xi_{\parallel} \gg \xi_{\perp}$, with $\chi_p(r)$ directly linked to $\xi_{\parallel}$. Thus, near-critical fluctuations occur within the solvation shell even when its width (set by $\xi_{\perp}$) is molecular in scale. 

To assess how these features mediate solute–solute interactions, we developed a morphometric model treating the solvation shell as a finite-width layer with internal pressure and effective surface tension, without molecular-level detail. Despite its simplicity, the model reproduces effective interaction potentials from molecular dynamics of mW and SPC/E water across solute sizes, interaction strengths, and thermodynamic states.

\begin{figure}[t]
    \centering
    \includegraphics{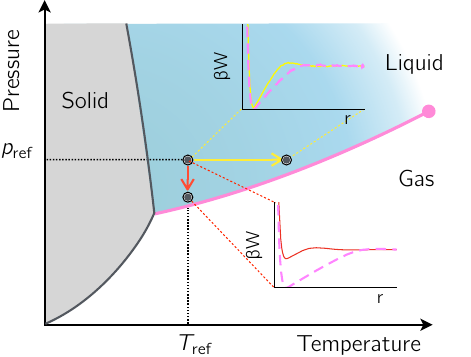}
    \caption{
Schematic phase diagram of water showing the thermodynamic routes used to evaluate solute–solute interactions. A representative isobar (yellow, $p_{\text{ref}}$) and isotherm (red, $T_{\text{ref}}$) move the system from the stable liquid toward the vapor–liquid coexistence line (pink) by (i) heating or (ii) lowering pressure. Near coexistence, surface criticality drives growth of the correlation length $\ell$, strengthening and extending hydrophobic attraction. Insets illustrate the effective interaction potential at the start (solid) and end (dashed) of each path. Bulk vapor–liquid criticality (pink circle) plays no role.}
\label{fig:diagram}
\end{figure}

The effective interaction between solutes is governed by two principal free energy contributions: a pressure oversaturation term and a dominant surface tension term. These parameters, though fitted to simulation data, are thermodynamically consistent and vary predictably with temperature and pressure. The solvation shell width, $\ell$, which corresponds to half the attractive range of the effective interaction potential, emerges as a central length scale. Its growth upon approaching vapor-liquid coexistence, whether by lowering pressure or raising temperature, serves as a clear indicator of emerging density correlations and near-(surface)-critical behaviour. Its dependence on thermodynamic fields and solute properties follows a scaling law derived from a binding potential theory for critical drying, enabling predictions for how the interaction changes with system parameters. Importantly, the scaling expression, in combination with the morphometric model, accounts for both the strengthening and longer range of hydrophobic attraction as the system approaches coexistence (as summarized in Fig.~\ref{fig:diagram}). We note that the solvation shell width -- and its pressure and temperature dependence -- has hardly been introduced or defined as a relevant quantity in prior studies of the hydrophobic effect.

We briefly clarify how this work connects to the LCW theory of hydrophobicity noted in Sec.~\ref{sec:intro}. LCW highlights key aspects of hydrophobic solvation: emergence of a liquid–vapor interface, the role of surface tension, length-scale crossover, and proximity to coexistence. However, it has limitations: it emphasizes changes in local density profiles with model or thermodynamic parameters but does not explain the fluctuation-driven origin of phenomena, make scaling predictions, or clarify factors shaping hydrophobic interaction potentials. Terms such as ``drying transition,''\cite{HuangChandler2000,HuangChandler2002} ``dewetting transition,''\cite{Chandler:2005aa,PatelGarde2012,Sarupria2009} and ``pre-transition effects''\cite{Katira2016} largely serve as shorthand for vapor-like regions near or between hydrophobes, without identifying the nature of the transition itself.  

Recasting the problem in terms of critical drying goes beyond LCW, placing it within established surface phase transition theory. This allows binding-potential and scaling methods to quantify near-critical fluctuations. Within this framework, the origin and non-Gaussian character of solvation-shell fluctuations, including transient vapor bubbles, follow naturally, as does the scaling of shell width with solute radius, affinity to solvent, pressure, and temperature. These elements underlie hydrophobic solvation and attraction, provide a mechanism for the inverse-temperature effect in extended solutes, and constrain the shape, range, and depth of effective interaction potentials.

Our work describes hydrophobic attraction in terms of solvation properties governed by surface phase behaviour, a framework that depends on proximity to coexistence and coarse-grained solute properties rather than fluid-specific chemistry. This shifts the focus from solvent structuring and hydrogen bonding to a surface phase transition as the origin of hydrophobic interactions, particularly for extended solutes. The approach is physically transparent and broadly applicable, reproducing the inverse temperature effect through the temperature dependence of $\ell$ and surface phase behaviour alone. Its validity is reinforced by the same effect appearing in a Lennard-Jones solvent, which lacks hydrogen bonding.  

Finally, we note that our framework describes a coarse-grained description for extended solutes. For very large ones, capillary evaporation may replace drying as the governing mechanism~\cite{Malijevski-Parry2015}. Exploring this crossover, particularly in biological systems, is a promising direction for future work.

\acknowledgements

The authors thank Bob Evans for many helpful conversations and a careful reading of the manuscript. The computer simulations were carried out using the computational facilities of the Advanced Computing Research Centre, University of Bristol, as well as the Isambard 2 UK National Tier-2 HPC Service (\href{http://gw4.ac.uk/isambard/}{http://gw4.ac.uk/isambard/}) operated by GW4 and the UK Met Office, and funded by EPSRC (EP/T022078/1). NBW acknowledges the support of the Leverhulme Trust grant RF-2024-173. \\

\appendix

\section{Molecular Dynamics simulations}
\label{app:mD}
All simulations were conducted using the LAMMPS~\cite{LAMMPS} MD engine within the isothermal-isobaric ensemble. Depending on the solute size and the type of measurement—whether for the profiles $\varrho(r)$ and $\chi_p(r)$ around an isolated solute or the effective potential $W(r)$ between a pair of solutes—we used periodic simulation boxes with linear dimensions of $L = (\SI{60}{\angstrom})^3$ and $L = (\SI{80}{\angstrom})^3$. We simulate systems with particle numbers ranging from $6920$ to $16200$, at a constant timestep of \SI{3}{\femto\second} using a Nose-Hoover style thermostat-barostat with a time constant \SI{500}{\femto\second} for both barostat and thermostat.

The spherical shell subvolumes necessary for computing $\langle \varrho(r)\rangle$ and $\langle V\varrho(r)\rangle$, which are required for determining $\varrho(r)$ and $\chi_p(r)$ as described above, were created using LAMMPS' ``chunks'' functionality. For illustration, the relevant portion of the LAMMPS input file is provided below.
\begin{widetext}
\begin{lstlisting}[breaklines]

# LAMMPS input file snippet to measure ingredients of \varrho(r) and \chi_p(r) across spherical shells

variable Rmaxrough equal 30
variable dr  equal 0.2
variable nsbin equal floor(${Rmaxrough}/${dr})
variable Rmax equal ${nsbin}*${dr}

compute chunks  all chunk/atom bin/sphere 0.0 0.0 0.0 0.0 ${Rmax} ${nsbin}  units box
variable counter atom 1
variable counter_volume atom vol*1 #couple the presence of a particle to volume fluctuation

compute n1      all  reduce/chunk chunks sum v_counter
compute n1a     all  chunk/spread/atom chunks c_n1
compute n1v     all  reduce/chunk chunks sum v_counter_volume
compute n1av    all  chunk/spread/atom chunks c_n1v
fix ave1 all ave/chunk 200 1 200 chunks c_n1a c_n1av norm all  ave running file shells-T${T}-P${P}-dr${dr}-Rs${solute_rad}.profile overwrite
\end{lstlisting}
\end{widetext}
\vspace*{5mm}

MD measurements of $W(r)$ were performed using the well-tempered metadynamics technique~\cite{barducci2008} as implemented in the collective variables package PLUMED~\cite{plumed}. This biasing scheme enhances the sampling of states away from the minimum of the effective potential. We identify the center-to-center distance between the two solute spheres $r$ as our collective variable and use the metadynamics algorithm to construct a history-dependent potential that is the accumulation of Gaussian kernels at times $n\tau$, deposited at discrete times $\tau$.

The history-dependent potential is the accumulated value

\begin{equation}
 V[r,t]= \sum_{n\tau<t} G(n\tau)\exp{
 \left[-\dfrac{\left(r-r(n\tau)\right)^2}{2\sigma^2}\right]}
\end{equation}
where the width $\sigma$ of the Gaussian is an input parameter and the height is chosen according to the well-tempered protocol\cite{barducci2008} as
\begin{equation}
G(n\tau)=G_0 \exp \left(-\frac{V[r (n\tau),n\tau]}{k_{B} \Delta T}\right),
\end{equation}
where $W_0$ is the initial height of the Gaussian and $\Delta T$ is obtained from the simulation temperature and the biasing factor $\gamma$ as $\Delta T = T (\gamma-1)$, where $\gamma$ is an input parameter. This ensures that over time the corrections gradually decrease in magnitude and the potential asymptotically converges to the potential of mean force $W(r)$ (plus a shifting constant) as

\begin{equation}
    V(r,t\rightarrow\infty)=-\dfrac{W(r)}{\gamma}+\mathrm{const.}
\end{equation}

We find that our simulations are stable and explore the free energy surface thoroughly with initial heights of $\SI{1}{\kilo\cal\per\mole}$, Gaussian widths of  $\SI{0.04}{\angstrom}$ with biases in the rage $5-10$ and accumulated every $\SI{500}{\femto\second}$. We initially equilibrate the water with the solutes {\em in situ} and let the pressure and temperature converge, with a warmup time of $\SI{5e6}{\femto\second}$. We then accumulate data every $\SI{500}{\femto\second}$ for a total of \SI{2e8}{\femto\second}.


\section{Monatomic water (mW) model} 
\label{app:mW}

Our MD simulations utilize a popular monatomic water (mW) model proposed by Molinero and Moore \cite{MolineroMoore2009}. This coarse-grained model represents a water molecule as a single particle and reproduces the tetrahedral network structure of liquid water using a parameterization of the Stillinger-Weber potential. Within the mW model, particles interact via the potential \cite{MolineroMoore2009}
\begin{widetext}
\begin{equation}
     \phi_{mw}(\mathbf{r}_i,\mathbf{r}_j,\mathbf{r}_k,\theta_{ijk}) = \sum_i\sum_{j>i}\phi_{mw,2}(\mathbf{r}_i,\mathbf{r}_j) \\
   +\sum_i\sum_{j\neq i}\sum_{k>j}\phi_{mw,3}(\mathbf{r}_i,\mathbf{r}_j,\mathbf{r}_k,\theta_{ijk})\:,   
\end{equation}
where the two-body, $\phi_{mw,2}$ and three-body, $\phi_{mw,3}$, potentials are
\begin{equation}
    \phi_{mw,2}(\mathbf{r}_i,\mathbf{r}_j) = \\A\varepsilon_{mw} \left[B\left(\frac{\sigma_{mw}}{r}\right)^4 - 1\right]\exp\left(\frac{\sigma_{mw}}{|\mathbf{r}_i-\mathbf{r}_j|-a\sigma_{mw}}\right)\;,
\end{equation}
\begin{align}
    \phi_{mw,3}(\mathbf{r}_i,\mathbf{r}_j,\mathbf{r}_k,\theta_{ijk}) &=  \lambda\varepsilon_{mw}\left[\cos\theta_{ijk} - \cos\theta_0\right]^2\exp\left(\frac{\gamma\sigma_{mw}}{|\mathbf{r}_i-\mathbf{r}_j|-a\sigma_{mw}}\right)\\
    \nonumber
    &\times\exp\left(\frac{\gamma\sigma_{mw}}{|\mathbf{r}_i-\mathbf{r}_k|-a\sigma_{mw}}\right)\;,
\end{align}
\end{widetext}
and $A = 7.049556277$, $B=0.6022245584$, $\gamma=1.2$ are constants which determine the form and scale of the potential, $\lambda=23.15$ is the tetrahedral parameter, $\theta_0=109.47^{\circ}$ is the angle favored between waters, $a=1.8$ sets the cut-off radius, $\sigma_{mw}=2.3925$\AA~is the diameter of a mw particle, and  $\varepsilon_{mw}=6.189\; \mathrm{kcal\;mol^{-1}}$ is the mW-mW (water-water) interaction strength. 

Our simulation box contains mW particles, with either a single spherical solute of radius $R_s$ centered at the origin or two identical solutes that can move relative to each other and interact through a purely repulsive Morse potential. The mW particles also interact with the solute via a Morse potential, with parameters set to make the interaction either purely repulsive or with a truncated  attractive component (see Appendix~\ref{app:interpot}).

\section{Solute-solute and solute-water interaction potential} 
\label{app:interpot}

We employ a smoothed Morse potential to describe the solute-solute and solute-water interactions. The Morse potential takes the form

\begin{equation}
\phi(r) = \varepsilon \left[ e^{-2\alpha(r-r_0)} - 2e^{-\alpha(r-r_0)} \right] \quad \text{for } r < r_c\:,
\end{equation}
where $\varepsilon$ is the well depth, $\alpha$ sets the lengthscale, $r_0$ sets the interaction distance, and $r_c$ is a cutoff parameter.  

We shift and smooth this bare Morse potential to ensure that both the potential and force vanishes  at the cutoff, as follows:

\begin{equation}
E(r) = \phi(r) - \phi(r_c) - (r - r_c) \left. \frac{d\phi}{dr} \right|_{r=r_c} \quad \text{for } r < r_c\:.
\end{equation}

To obtain purely repulsive solute-solute and solute-water interactions we set $\varepsilon=1.0, \alpha=3.0, r_c=r_0$, where $r_0=R_s$ for solute-water interactions and $r_0=2R_s$ for solute-solute interactions.

For the simulations of the solute having $R_s=10\,${\AA} attractive solute-water interactions, we set $r_c=15\,${\AA} and varied $\varepsilon=\varepsilon_{sw}$ in the range $0.2,0.4.0.6\,$kcal/mol.\\

\section{The local density profile and its pressure derivatives}
\label{app:chip}

For a spherical solute centered at the origin, we calculate the radial profiles $\varrho(r), \chi_p(r)$ and $\upsilon_p(r)$ by forming spherical shell subvolumes of thickness $\delta r$ and volume $\delta V=4\pi r^2\delta r$ as described elsewhere~\cite{Wilding:2024}. 
Within each subvolume, we calculate the local density as a histogram obtained by binning $M$ independent measurements of the solvent particle positions into the set of subvolumes~\cite{EvansWilding2015,EvansStewartWilding2017} and calculating

\begin{equation}
 \varrho(r)\equiv\langle \rho(r)\rangle =\frac{1}{M}\sum_{i=1}^M\rho_i(r)\;,
\end{equation}
where $\rho_i=N_i(r)/\delta V$, with $N_i$ the instantaneous value of the particle number 
in the subvolume that extends from $r$ to $r+\delta r$ having volume $\delta V$.   $\varrho(r)$ has units $m^{-3}$.

The local compressibility provides a sensitive measure of enhancement in local density fluctuations \cite{EvansStewart2015,Eckert_2023,Wilding:2024}. In the constant-$NpT$ ensemble, the appropriate definition is 

\begin{equation}
\chi_p(r) \equiv \frac{\partial  \varrho(r)}{\partial p}\bigg|_T   \:,
\end{equation}
which has units of $J^{-1}$.

For measurements, we express $\chi_p(r)$ as a covariance relation. This relation is derived from the isothermal-isobaric partition function for an inhomogeneous system as follows

\begin{widetext}
\begin{equation}
Z= \int_0^\infty dV \prod_{i=1}^N \int d{\bf r_i}\exp{\left(-\beta[E(\{ {\bf r_i} \}^{N})+pV +\int drV_{ext}(r)\varrho(r)]\right)},
\end{equation}
\end{widetext}
with $\beta=1/k_BT$ and $V_{ext}$ is the radial potential exerted by the solute on the solvent.  

The corresponding free energy is
\begin{equation}
{\cal G}=-\beta^{-1}\ln Z,
\end{equation}
and the local density profile $\varrho(r)$ is given by the (functional) derivative 

\begin{equation}
   \frac{\delta {\cal G}}{\delta V_{ext}}=\varrho(r) \:.
\end{equation}

Utilizing the definition of local isothermal compressibility one finds
\begin{widetext}
\begin{eqnarray}
   \chi_p(r)&\equiv&\frac{\partial \varrho(r)}{\partial p}\bigg|_T=\frac{\partial^2 {\cal G}}{\delta V_{ext}(r)\partial p}\label{eq:chidef}=-\beta^{-1}\frac{\partial^2 \ln Z}{\delta V_{ext}(r)\partial p}\\
   &=& -\frac{1}{\beta Z}\left(\frac{\partial^{2}Z}{\partial  V_{ext}(r)\partial p}\right) +\frac{1}{\beta Z^2}\left(\frac{\partial Z}{\partial  V_{ext}(r)}\right)\left(\frac{\partial Z}{\partial  p}\right)\\
      &=&-\beta\left( \langle V\rho(r)\rangle- \langle \rho(r)\rangle\langle V\rangle\right)\\
   \beta^{-1}\chi_p(r)&=&-\left( \langle V\rho(r)\rangle- \langle \rho(r)\rangle\langle V\rangle\right),
\end{eqnarray}
\end{widetext}
 where the minus sign is indicative of an anticorrelation between $V$ and $\varrho (r)$.

 Note that in the bulk $\chi_p(r)$ decays to $\varrho_b\kappa_T$, where $\varrho_b$ is the bulk number density and

\begin{equation}
    \kappa_T\equiv -\frac{1}{V}\frac{\partial V}{\partial p}=\frac{1}{k_BT} \frac{\langle V^2\rangle-\langle V\rangle^2}{\langle V\rangle}\:,
    \label{eq:bcomp}
\end{equation}
is the bulk isothermal compressibility.

Now let us define a higher-order derivative of the free energy with respect to pressure:

\begin{widetext}
\begin{equation}
    \upsilon(r)\equiv \frac{\partial^2 \varrho(r)}{\partial p^2}=\frac{\partial^3 {\cal G}}{\delta V_{ext}(r)\partial p^2}=-\beta\frac{\partial^3 \ln Z}{\delta V_{ext}(r)\partial p^2},
\end{equation}
which has units of $m^3 J^{-2}$.
This yields

\begin{eqnarray}
 \upsilon(r)&=& -\frac{1}{\beta Z}\frac{\partial^{3}Z}{\delta V_{ext}(r)\partial p^{2}}
+\frac{2}{\beta Z^2}\left(\frac{\partial^{2}Z}{\delta  V_{ext}(r)\partial p}\right) \left(\frac{\partial Z}{\partial p}\right)\\\nonumber
&\:& -\frac{2}{\beta Z^3} \left(\frac{\partial Z }{\delta  V_{ext}(r)}\right) \left(\frac{\partial Z}{\partial p}\right)^{2}
+\frac{1}{\beta Z^2}\left(\frac{\partial Z}{\delta  V_{ext}(r)}\right) \left(\frac{\partial^{2}Z}{\partial p^{2}}\right),
\end{eqnarray}
which in terms of observables, takes the form:

\begin{equation}
\upsilon (r)= (k_BT)^{-2}\left[\langle V^2\rho(r)\rangle -2\langle \rho(r)V\rangle\langle V\rangle+2\langle \rho(r)\rangle\langle V\rangle^2 -\langle\rho(r)\rangle\langle V^2\rangle\right]\:.
\label{eq:upsilon}
\end{equation}
\end{widetext}

Note, however, that as a higher-order derivative, $\upsilon_p(r)$ is inherently more difficult to measure accurately than $\chi_p(r)$. The left-hand side of the covariance relation Eq.\,(\ref{eq:upsilon}) is small in magnitude compared to the individual summation and subtraction terms on the right-hand side, necessitating extremely precise statistical sampling to obtain a reliable signal.\\

\section{Selected bulk properties of MW model} 
\label{app:bulkprops}
The temperature dependence of the vapor-liquid phase behavior and the vapor-liquid surface tension for the pure MW fluid has been previously calculated~\cite{CoeEvansWildingJCP2022}. In Tab.~\ref{tab:tabbulk}, we present results for the bulk liquid number density and isothermal compressibility (computed using Eq.\,(\ref{eq:bcomp})) at the state points investigated in this study. The uncertainties in $\varrho_b$ are typically $0.001 \times 10^{28}$ m$^3$, while those in $\kappa_T$ are generally $0.003 \times 10^{-10}$ m$^3$J$^{-1}$.

\begin{table}[t!]
\centering
\caption{Measured values of bulk number density and isothermal compressibility of the mW model at various temperatures and pressures. }
  \renewcommand{\arraystretch}{0.7} 
\begin{tabular}{lr|rr}
$T$ (K) & $p$ (atm) & $\varrho_b/10^{28}$ (m$^{-3}$) & $\kappa_T/10^{-10} ($m$^{3}$J$^{-1}$) \\
\midrule

300 & 0 & 3.333 & 1.967\\
325 & 0 & 3.314 & 1.991\\
350 & 0 & 3.291 & 2.072\\
375 & 0 & 3.265 & 2.180\\
400 & 0 & 3.237 & 2.350\\
425 & 0 & 3.207 & 2.536\\
426 & 0 & 3.205 & 2.540 \\
300 & 100 & 3.340 & 1.920\\
300 & 200 & 3.346 & 1.892 \\
300 & 300 & 3.353 & 1.843\\
300 & 500 & 3.365 & 1.759\\ 
300 & 750 & 3.379 & 1.676 \\
300 & 1000 & 3.393 & 1.587 \\
300 & 1500 & 3.419 & 1.456 \\
300 & 2000 & 3.444 & 1.343\\
300 & 3000 & 3.487 & 1.168\\
300 & 4000 & 3.526 & 1.030\\
300 & 5000 & 3.561 & 0.927\\
426 & 500 & 3.244 & 2.174\\
426 & 750 &  3.261 & 2.205\\
426 & 1000 & 3.277 & 1.920\\
426 & 1500 & 3.308 & 1.726\\
\bottomrule
\end{tabular}
\label{tab:tabbulk}
\end{table}


\bibliography{references.bib}

\end{document}